\documentclass{emulateapj}
\usepackage{amssymb}
\usepackage{amsmath}
\usepackage{hyperref}

\newcommand{\mbh}{\ensuremath{M_{\rm{BH}}}\,}
\newcommand{\er}{\ensuremath{\lambda_{\rm Edd}\,}}
\newcommand{\Lx}{\ensuremath{L_{\rm{[2-10keV]}}}}

\newcommand{\rev}[1]{{ #1}}

\received{August 10, 2018}
\revised{September 28, 2018}
\accepted{October 4, 2018}

\shorttitle{FMOS Survey of broad-line AGN in deep survey fields}
\shortauthors{Schulze et al.}

\begin{document}

\title{An FMOS Survey of moderate-luminosity broad-line AGN in COSMOS, SXDS and E-CDF-S}

\author{Andreas Schulze\altaffilmark{1,16}, John D. Silverman\altaffilmark{2}, Daichi Kashino\altaffilmark{3}, Masayuki Akiyama\altaffilmark{4}, Malte Schramm\altaffilmark{1}, Dave Sanders\altaffilmark{5}, Jeyhan Kartaltepe\altaffilmark{6}, Emanuele Daddi\altaffilmark{7}, Giulia Rodighiero\altaffilmark{8}, Alvio Renzini\altaffilmark{9}, Nobuo Arimoto\altaffilmark{10}, Tohru Nagao\altaffilmark{11}, Annagrazia Puglisi\altaffilmark{7,9}, Benny Trakhtenbrot\altaffilmark{3,12,17}, Francesca Civano\altaffilmark{13,14}, Hyewon Suh\altaffilmark{15}}
\email{E-mail: andreas.schulze@nao.ac.jp}

\altaffiltext{1}{National Astronomical Observatory of Japan, Mitaka, Tokyo 181-8588, Japan}
\altaffiltext{2}{Kavli Institute for the Physics and Mathematics of the Universe (WPI), The University of Tokyo, Kashiwa, Chiba 277-8583, Japan}
\altaffiltext{3}{Department of Physics, ETH Z{\" u}rich, Wolfgang-Pauli-strasse 27, CH-8093, Z{\" u}rich, Switzerland}
\altaffiltext{4}{Astronomical Institute, Tohoku University, Aramaki, Aoba-ku, Sendai, 980-8578, Japan}
\altaffiltext{5}{Institute for Astronomy, University of Hawaii, 2680 Woodlawn Drive, Honolulu, HI, 96822, USA}
\altaffiltext{6}{School of Physics and Astronomy, Rochester Institute of Technology, 84 Lomb Memorial Drive, Rochester, NY 14623, USA}
\altaffiltext{7}{CEA, IRFU, DAp, AIM, Universit\'{e} Paris-Saclay, Universit\'{e} Paris Diderot, Sorbonne Paris Cit\'{e}, CNRS, F-91191 Gif-sur-Yvette, France}
\altaffiltext{8}{Dipartimento di Fisica e Astronomia, Universita di Padova, vicolo Osservatorio, 3, I-35122, Padova, Italy}
\altaffiltext{9}{Instituto Nazionale de Astrofisica, Osservatorio Astronomico di Padova, vicolo dell' Osservatorio 5, I-35122, Padova, Italy}
\altaffiltext{10}{Astronomy Program, Department of Physics and Astronomy, Seoul National University, 599 Gwanak-ro, Gwanaku-gu, Seoul 151-742, Korea}
\altaffiltext{11}{Graduate School of Science and Engineering, Ehime University, 2-5 Bunkyo-cho, Matsuyama 790-8577, Japan}
\altaffiltext{12}{School of Physics and Astronomy, Tel Aviv University, Tel Aviv 69978, Israel}
\altaffiltext{13}{Harvard-Smithsonian Center for Astrophysics, 60 Garden Street, Cambridge, MA 02138, USA}
\altaffiltext{14}{Yale Center for Astronomy and Astrophysics, 260 Whitney Avenue, New Haven, CT 06520, USA}
\altaffiltext{15}{Subaru Telescope, 650 North A'ohoku Place, Hilo, Hawaii, 96720, USA}
\altaffiltext{16}{EACOA Fellow}
\altaffiltext{17}{Zwicky Fellow}



\begin{abstract}
We present near-IR spectroscopy in $J$- and $H$-band for a large sample of 243 X-ray selected moderate-luminosity type-1 AGN in the COSMOS, SXDS and E-CDF-S survey fields using the multi-object spectrograph Subaru/FMOS. Our sample covers the redshift range $0.5\lesssim z\lesssim3.0$ and an X-ray luminosity range of $10^{43}\lesssim \Lx \lesssim 10^{45}$~erg s$^{-1}$. We provide emission-line properties and derived virial black hole mass estimates, bolometric luminosities and Eddington ratios, based on H$\alpha$ (211), H$\beta$ (63) and \ion{Mg}{2} (4). We compare line widths, luminosities and black hole mass estimates from H$\alpha$ and H$\beta$ and augment these with commensurate measurements of \ion{Mg}{2} and \ion{C}{4} detected in optical spectra. We demonstrate the robustness of using H$\alpha$, H$\beta$ and \ion{Mg}{2} as reliable black hole mass estimators for high-$z$ moderate-luminosity AGN, while the use of \ion{C}{4} is prone to large uncertainties ($\gtrsim0.4$~dex). We extend a recently proposed correction based on the \ion{C}{4} blueshift to lower luminosities and black hole masses. While our sample shows an improvement in their \ion{C}{4} black hole mass estimates, the deficit of high blueshift sources reduces its overall importance for moderate-luminosity AGN, compared to the most luminous quasars.
In addition, we revisit luminosity correlations between $L_{\rm{bol}}$, \Lx, $L_{\rm{[OIII]}}$, $L_{5100}$ and $L_{\rm{H}\alpha}$ and find them to be consistent with a simple empirical model, based on a small number of well-established scaling relations. We finally highlight our highest redshift AGN, CID 781, at $z=4.6$ which shows the lowest black hole mass  ($\sim10^8$M$_\odot$) among current near-IR samples at this redshift, and is in a state of fast growth.
\end{abstract}




\section{Introduction}
The study of the population of active galactic nuclei (AGN) and its evolution out to high redshift is closely entangled with the understanding of galaxy evolution and the role supermassive black holes (SMBH) play therein. Evidence for a link between SMBH growth and galaxy evolution comes from observational results \citep[e.g.][]{Magorrian:1998,Silverman:2008,Alexander:2012,Fabian:2012,Kormendy:2013} and from theoretical arguments \citep[e.g.][]{Silk:1998}. It is also an essential ingredient in numerical simulations and semi-analytical models \citep[e.g.][]{DiMatteo:2005,Somerville:2008,Vogelsberger:2014,Schaye:2015}. However, the details and especially the causal connection between the two remain poorly understood. An essential cosmic period for studying the connection between black hole growth and star formation/galaxy evolution is the redshift range $1<z<3$, corresponding to the peak epoch of star formation and black hole activity and the beginning of their decease \citep{Boyle:1998,Silverman:2008,Aird:2015}.

Disentangling the black hole growth history requires knowledge of not only the AGN luminosities but also their black hole masses and normalized accretion rates, i.e. Eddington ratios. Such knowledge allows a more meaningful study of the demographics and cosmic evolution of the AGN population \citep[e.g.][]{McLure:2004,Netzer:2007b,Vestergaard:2009,Trakhtenbrot:2012,Kelly:2013,Schulze:2015}. 
These studies revealed an intrinsically broad distribution of Eddington ratios, with an upper boundary around the Eddington limit and with the mean Eddington ratio increasing with increasing redshift. They also demonstrated the role of black hole mass in the downsizing trends seen in the AGN luminosity function \citep{Schulze:2010,Schulze:2015}. Furthermore, knowledge of the the SMBH mass is essential to study the cosmic evolution of the scaling relation between SMBH mass and its host galaxy properties out to high redshift \citep[e.g.][]{Peng:2006,Salviander:2007,Park:2015}.

For broad-line (type-1) AGN SMBH mass estimates can be obtained from single-epoch spectroscopy using the so-called virial method \citep[e.g.][]{McLure:2002,Shen:2013}. 
Under the assumption that the motion of the broad-line region (BLR) gas is virialised, it is possible to infer the central black hole mass. This became feasible thanks to extensive reverberation mapping campaigns \citep{Peterson:2004} which established an empirical scaling of the BLR size with AGN continuum luminosity \citep{Kaspi:2000,Bentz:2009}. The velocity of the BLR gas can be inferred from the width of the broad emission lines. The virial method builds on these reverberation mapping results and is directly calibrated to it, at least for the broad H$\beta$ line \citep{Vestergaard:2006}. Other broad emission lines, like H$\alpha$, \ion{Mg}{2} and \ion{C}{4} are commonly calibrated to H$\beta$ \citep[e.g.][]{Greene:2005,McGill:2008,Park:2017}. More recent reverberation mapping experiments targeting such lines generally confirm the validity of these calibrations \citep{Grier:2017,Lira:2018}.
At $z>1$ H$\beta$ moves out of the optical spectral range, which necessitates either the use of a different broad-line to infer the black hole mass or observations in the near-IR.

Since the broad Balmer lines are considered to provide the most reliable black hole mass estimates, observations of high-$z$ broad-line AGN in the near-IR, covering the rest-frame optical, are highly valuable. Furthermore, the rest-frame optical range, easily observed at low redshift, provides additional information on the AGN structure and demographics, from probing the outer accretion disk emission, the optical BLR and the narrow line region (NLR). A good tracer of the NLR, its size and its kinematics, largely due to its strength, is the [\ion{O}{3}]$\lambda\lambda4959,5007$\AA\, line doublet \citep[e.g.][]{Stern:2013,Mullaney:2013,Shen:2014}.

These considerations  have provided the motivation for studies of broad-line AGN in the near-IR using either single-object \citep{Yuan:2003,Shemmer:2004,Sulentic:2004,Sulentic:2006,Netzer:2007,Dietrich:2009,Marziani:2009,Greene:2010,Shen:2012b,Ho:2012,Bongiorno:2014,Collinson:2015,Jun:2015,Zuo:2015,Mejia:2016,Trakhtenbrot:2016,Coatman:2016,Saito:2016,Jun:2017,Bisogni:2017,Vietri:2018} or, more recently, multi-object spectroscopy with instruments such as FMOS \citep{Nobuta:2012,Matsuoka:2013,Suh:2015,Karouzos:2015}.

Most of such studies, and particularly the earlier ones, focused on extremely luminous sources, powered by SMBHs with high Eddington ratio (0.1-1.0) and/or high SMBH mass, reaching $\sim10^{10} M_\odot$. Such very luminous systems are intrinsically rare and do not represent the \emph{typical} AGN population at $z>1$ \citep{Richards:2006b,Ross:2013,Aird:2015}.

In particular, moderate-luminosity AGN ($10^{44}\lesssim L_{\rm bol}\lesssim 10^{46}$) at high redshift ($z>1$) are important probes of black hole growth and AGN physics since they trace the bulk of the population and have direct analogs (e.g. similar luminosity) at lower $z$. Deep X-ray surveys are very effective at detecting this population of unobscured to moderately obscured AGN \citep[e.g.][]{Brandt:2015, Xue:2017}. The COSMOS field \citep{Scoville:2007} has been very important in this respect, due to its deep X-ray coverage \citep{Civano:2016} and available multi-wavelength data \citep[e.g.][]{Laigle:2016,Smolcic:2017}. Another important X-ray extragalactic survey field is the Subaru-XMM-Newton Deep Survey \citep[SXDS;][]{Ueda:2008,Furusawa:2008,Akiyama:2015}. Both fields constitute the UltraDeep layer of the ongoing Hyper Suprime-Cam Subaru Strategic Program (HSC-SSP) survey  on the Subaru telescope \citep{Aihara:2018}, where they will be covered in $grizy$ to a depth of $i\simeq28$~mag. 

Additional information on the AGN population (i.e., black hole masses and Eddington ratios) is important for a broad range of applications, including the investigation of black hole-galaxy coevolution \citep{Jahnke:2009,Merloni:2010,Cisternas:2011,Schramm:2013,Sun:2015}, the connection between AGN activity and host star formation \citep{Rosario:2013}, the study of AGN demographics itself \citep{Trump:2009,Nobuta:2012,Schulze:2015} or the dependence of observed AGN properties on black hole mass and Eddington ratio \citep{Brightman:2013}.

In this work, we present near-IR spectroscopy, covering the broad H$\alpha$, H$\beta$ and \ion{Mg}{2} lines, for an unprecedentedly large sample of 243 moderate-luminosity AGN at $0.5\lesssim z\lesssim3.0$ obtained with the Fibre Multi-Object Spectrograph \citep[FMOS;][]{Kimura:2010} on the Subaru telescope. Our main target fields are COSMOS and SXDS. We augment the sample by additional observations in the Extended Chandra Deep Field South \citep[E-CDF-S;][]{Lehmer:2005}. Initial results for a subset of 43 AGN of our sample in COSMOS and E-CDF-S have been presented in \citet[][hereafter M13]{Matsuoka:2013}. For the majority of our sample in SXDS broad H$\alpha$ measurements have been presented in \citet[][hereafter N12]{Nobuta:2012}, but they did not provide direct black hole mass estimates from that line. Here we provide those estimates as well as broad H$\beta$ detections for these and additional broad-line AGN in SXDS. A complete black hole mass catalog of AGN in COSMOS, including results from optical spectroscopy, will be presented in a future publication.

We present the sample and the observations in section~\ref{sec:sample}. In section~\ref{sec:analysis} we present our spectral measurements and an assessment of the prominence of reddening in our sample. In section~\ref{sec:results} we present the derived estimates of SMBH mass and Eddington ratio and discuss correlations between different virial SMBH mass estimators. In section~\ref{sec:discuss} we give a discussion of  topics pertaining to our near-IR spectra, including the virial SMBH mass estimators, comparison with low redshift analogs of similar luminosity, correlations between different bolometric luminosity indicators and the early growth of the black hole population. We present our conclusions in section~\ref{sec:conclu}.
Throughout this paper we use a Hubble constant of $H_0 = 70$ km s$^{-1}$ Mpc$^{-1}$ and cosmological density parameters $\Omega_\mathrm{m} = 0.3$ and $\Omega_\Lambda = 0.7$. Magnitudes are expressed in the AB system.

\section{Sample and Observations} \label{sec:sample}
FMOS is a near-infrared fiber spectrograph mounted on the Subaru telescope\footnote{The instrument was decommissioned in 2016}. It allows placement of  400 fibers ($1.2\arcsec$ diameter each) over a circular region of $30\arcmin$ diameter. Spectra for 200 targets can be obtained simultaneously over the field of view in cross-beam switching mode, which dithers targets between close fiber pairs for improved sky subtraction. In addition, an OH-airglow suppression filter \citep{Maihara:1994,Iwamuro:2001} masks out regions of strong atmospheric emission lines in $J$-band and $H$-band. Observations can be carried out in two resolution modes. The low resolution mode (LR)  allows simultaneous coverage of J-band and H-band over the wavelength range $0.9-1.8~\mu m$ at a spectral resolution of $\lambda / \Delta \lambda \approx600$. The high resolution mode (HR) achieves a spectral resolution of  $\lambda /\Delta \lambda \approx2600$, but requires four high-resolution gratings to fully cover the $J$ and $H$-band.

The FMOS survey we are using in this work consists of four observational efforts carried out with different spectral resolutions. In the COSMOS field we have conducted a survey in LR mode (Kartaltepe et al., in prep) and a survey in HR mode \citep{Silverman:2015}. Both surveys cover  the full 2~deg$^2$ area of the COSMOS field. In SXDS and E-CDF-S FMOS we carried out observations in LR mode only. We here present and combine results obtained from all four efforts. 

\subsection{Target selection}
The FMOS survey in COSMOS was primarily designed as a near-IR survey of high$-z$ galaxies, in particular targeting optical/near-IR selected star forming galaxies  at $z\sim1.6$ \citep{Kashino:2013,Zahid:2014,Silverman:2015,Kashino:2017a,Kashino:2017b} and far-IR and mid-IR sources using {\it Herschel}/PACS or {\it Spitzer}/MIPS  \citep{Kartaltepe:2015,Puglisi:2017}. In addition, the survey targeted X-ray selected AGN, which are the focus of the present work. Initial results on the LR AGN sample have been presented in M13 and \citet{Brightman:2013}. 

The AGN selection is based on the X-ray point source catalogs from \textit{Chandra} \citep{Elvis:2009,Civano:2012,Civano:2016} and \textit{XMM-Netwon} \citep[XMM-COSMOS;][]{Cappelluti:2009,Brusa:2010}, with sensitivities
of $f_{[0.5-2.0~\rm{keV}]}>2\times10^{-16}$~erg~cm$^{-2}$ s$^{-1}$  and $f_{[2-10~\rm{keV}]}>7.3\times10^{-16}$~erg~cm$^{-2}$ s$^{-1}$   in the soft and hard band, respectively for Chandra and 
$f_{[0.5-2.0~\rm{keV}]}>5\times10^{-16}$~erg~cm$^{-2}$ s$^{-1}$ and $f_{[2-10~\rm{keV}]}>3\times10^{-15}$~erg~cm$^{-2}$ s$^{-1}$ for  \textit{XMM-Netwon}. Out of these catalogs we targeted both type-1 and type-2 AGN with spectroscopic or photometric redshifts\footnote{We make use of photometric redshifts for part of the type-2 AGN population and use spectroscopic redshifts whenever available} allowing the detection of either H$\alpha$, H$\beta$ or \ion{Mg}{2} within the spectral coverage. Over the wavelength range covered by FMOS, H$\alpha$ can be detected between $0.5<z<1.7$, H$\beta$ between $1.2<z<2.7$ and \ion{Mg}{2} between $2.8<z<5.3$, with gaps at $z\sim1.1$, $\sim1.8$ and $\sim4.0$ respectively, due to strong atmospheric absorption between $J$ and $H$ band. The aim of the FMOS NIR campaign was to target the H$\alpha$ or H$\beta$ lines. The \ion{Mg}{2} targets were included as fillers.

For the LR observations we used the XMM-COSMOS \citep{Brusa:2010} and C-COSMOS \citep{Civano:2012} catalogs as input. For the HR campaign, we used the C-COSMOS catalog for the central square degree observations, augmented by the {\it Chandra} COSMOS-Legacy survey \citep{Civano:2016,Marchesi:2016} for the outer area.
We purposefully re-observed AGNs in HR mode that already had LR data, since the HR mode has a higher throughput, to investigate the impact of spectral resolution on our results.
In addition to AGN specifically targeted as X-ray sources, some are also targeted by other selection criteria, e.g. as {\it Herschel} sources. We include all FMOS spectra of X-ray detected AGN, irrespective of their initial selection. Furthermore we note that all detected broad-line AGN in our sample are included in the {\it Chandra} COSMOS-Legacy survey \citep{Marchesi:2016}. We associate each of our FMOS COSMOS AGN to the respective X-ray source and optical counterpart given in \citet{Marchesi:2016}. 
While the full AGN sample contains both type-1 and type-2 AGN, in this paper we only focus on the type-1, i.e. broad-line, AGN sample. 
We targeted broad-line AGN down to a limiting magnitude of $J_{\rm{AB}} = 23$, with preference given to those at $J_{\rm{AB}} < 21.5$. 
Multi-wavelength photometry is provided in \citet{Laigle:2016}.

For the SXDS field, the FMOS observations focused on spectroscopic follow up of X-ray sources detected with \textit{XMM-Newton} \citep{Ueda:2008}. The X-ray observations cover a 1.3 deg$^2$ field with a flux sensitivity of $f_{[0.5-2.0~\rm{keV}]}\sim1\times10^{-15}$~erg~cm$^{-2}$ s$^{-1}$ and $f_{[2-10~\rm{keV}]}\sim3\times10^{-15}$~erg~cm$^{-2}$ s$^{-1}$.
Optical counterparts have been presented in \citet{Akiyama:2015} and additional multi-wavelength photometry, including Subaru/HSC, for SXDS sources is provided in \citet{Mehta:2018}. 

The AGN sample in E-CDF-S \citep{Lehmer:2005} is similar to that presented in M13. We targeted X-ray AGN detected in the central 4Ms area\footnote{The deeper 7Ms data was not available when conducting the survey} \citep{Luo:2010,Xue:2011}. Priority was given to those with $R_{\rm{AB}} < 22$. We associate optical/near-IR counterparts based on the multi-wavelength catalog by \citet{Hsu:2014}.


We took advantage of the high multiplex factor of the FMOS spectrograph to target a large number of both type-1 and type-2 AGN in each of the campaigns. For the COSMOS-LR campaign we observed in total 932 AGN in the COSMOS {\it Chandra} Legacy catalog. In the COSMOS-HR campaigns 892 targets were observed. In the SXDS field 851 out of the full sample of 896 unique AGN candidates in SXDS have been targeted with FMOS \citep{Akiyama:2015}. 
We here only consider those AGN for which we robustly detect a broad emission line in the FMOS spectrum, as listed in Table~\ref{tab:sample}. A study of the type-2 population will be presented in Kashino et al. (in prep).

\rev{For the fields of COSMOS and E-CDF-S, almost all of the type-1 AGN considered here had prior spectroscopic redshifts from optical spectroscopy obtained from several observing programs in COSMOS \citep{Trump:2007,Lilly:2007,Coil:2011,Alam:2015,Hasinger:2018} and E-CDF-S \citep{Szokoly:2004,Popesso:2009,Silverman:2010}, as listed in \citet{Marchesi:2016} and \citet{Xue:2016}. Two targets only had a photo-$z$ \citep{Salvato:2011} before the FMOS observations. In SXDS the FMOS observations have been carried out for identification and redshift determination in parallel to optical spectroscopy, as discussed in \citet{Akiyama:2015}.}

\subsection{Observations and Data Reduction}
We provide a brief summary of the observations and data reduction. Further details can be found in M13 and \citet{Akiyama:2015} for the LR survey and in \citet{Silverman:2015} for the HR survey. Observations for the LR survey over the full COSMOS field were obtained between 2010-2012 (semesters S10B-S12A). Total exposure times are around 2-3.5 hours on-source per pointing. The HR survey in COSMOS was conducted between 2012-2016, with observations of the central square degree carried out in 2012-2014 and of the outer area in 2015-2016. Several gratings have been used (H-LONG, J-LONG, H-SHORT, H-SHORT-prime), especially to cover both the H$\alpha$ and H$\beta$ region for $z\sim1.6$ galaxies and AGN. Total on-source exposure times are 3-5 hours per observation. 
Observations in SXDS have been carried out between 2009-2011 in guaranteed, engineering, and open-use time over 22 pointings. Typical exposure times per pointing range between 1-5 hours.
Observations in both LR and HR mode were carried out in cross-beam switching mode. 
We performed data reduction, wavelength and flux calibration using the publicly available pipeline FIBRE-pac \citep[FMOS Image-Based REduction package;][]{Iwamuro:2012}, providing 1D and 2D reduced spectra and their error spectra. 

While an initial flux calibration is performed during the data reduction process using spectra of bright stars, we scale the flux of each spectrum to its H-band or J-band magnitude to account for flux loss due to aperture effects, variable seeing and other factors.
For COSMOS  we are using the deep near-IR photometry from UltraVISTA \citep{McCracken:2012}, taken from the COSMOS2015 catalog \citep{Laigle:2016}. The near-IR photometry for the SXDS field is taken from the VISTA Deep Extragalactic Observations (VIDEO) survey \citep{Jarvis:2013} as provided in the recent SPLASH-SXDF catalog \citep{Mehta:2018}. For both surveys, we use VISTA $J$ or $H$ magnitudes measured within a 3\arcsec\ aperture. 
For the E-CDF-S we are using the near-IR photometry taken from VIDEO DR3 \citep{Jarvis:2013}, with the VISTA $J$ or $H$ magnitudes corrected to an aperture of 2\arcsec\ diameter.
This approach to absolute flux calibration ignores the effect of AGN variability \citep[e.g.][]{VandenBerk:2004,Caplar:2017}, which is expected to be of the order of 0.2~mag for our sample, based on a comparison of near-IR photometry between VIDEO and the Ultra Deep Survey (UDS) in the SPLASH-SXDF catalog.

\begin{figure}
\centering
\resizebox{\hsize}{!}{ \includegraphics[clip]{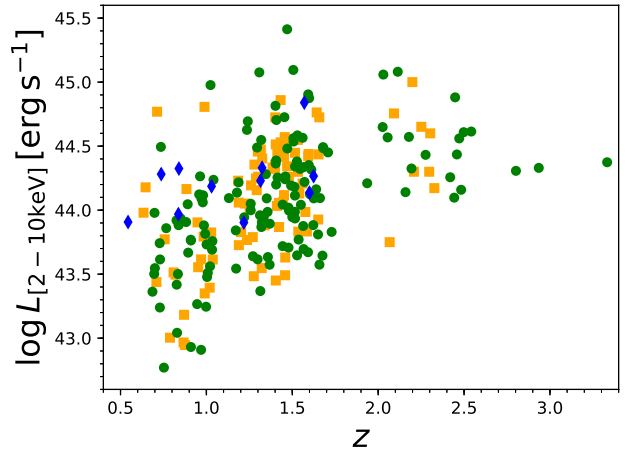} }
\caption{Distribution of redshift vs. 
2-10~keV X-ray luminosity for the X-ray selected broad-line AGN FMOS sample. Green squares denote AGN in COSMOS, orange squares are for AGN in SXDS and blue diamonds mark AGN in E-CDF-S. We use the same color scheme to distinguish the three survey fields throughout this paper. We have omitted CID~781 at $z=4.6$ from the plot.}
\label{fig:zlum} 
\end{figure}

\begin{deluxetable}{l  cccc}
\tabletypesize{\scriptsize}
\tablecaption{Sample overview}
\tablewidth{8cm}
\tablehead{ \colhead{Sample} & \colhead{Total} & \colhead{H$\alpha$} & \colhead{H$\beta$} & \colhead{\ion{Mg}{2}} 
}
\startdata
COSMOS &  145 & 121 & 39 & 4 \\  \noalign{\smallskip}
SXDS & 87 & 79 & 24 & 0 \\ \noalign{\smallskip}
E-CDF-S &  11 & 11 & 0 & 0 \\ \noalign{\smallskip}
Total &  243 & 211 & 63 & 4 \\
\enddata
\label{tab:sample}
\end{deluxetable}

\subsection{Sample properties}
Of all targets observed with FMOS, we include those that have a detected broad emission line with a full width half maximum FWHM$>1000$~km s$^{-1}$ in either H$\alpha$, H$\beta$ or \ion{Mg}{2}. We initially fit a parent sample of FMOS AGN spectra with sufficient signal to noise (S/N) as described in section~\ref{sec:line} using both a spectral model which only includes narrow emission lines and one with the addition of a broad-line. We classify the spectrum as a broad-line AGN based on the FMOS spectra if the addition of a broad component leads to a significant improvement of the best-fit in terms of reduced $\chi^2$.

We provide the sample statistics for the three surveys in Table~\ref{tab:sample}. In total we include 243 objects, with 145 in COSMOS, 87 in SXDS and 11 in E-CDF-S. Broad H$\alpha$ is detected for most of them (211), while  \ion{Mg}{2} is only detected in four cases at $z>2.8$. Our highest redshift AGN is CID~781 at $z=4.6$. \rev{Our sample provides H$\alpha$ measurements for spectroscopically confirmed type-1 AGN over the respective redshift range for $\sim 53$\% of the objects in the COSMOS-Legacy survey down to $J<21.5$~mag.}

We show the redshift and X-ray luminosity \Lx\ distribution for our sample in Figure~\ref{fig:zlum}. The X-ray luminosities are absorption corrected assuming an X-ray spectral index $\Gamma=1.8$, taken from \citet{Marchesi:2016} for COSMOS and  \citet{Akiyama:2015} for SXDS. For E-CDF-S we adopt the absorption corrected $L_{[0.5-7.0\rm{keV}]}$ luminosity from either the 7Ms CDFS catalog by \citet{Luo:2017} or the E-CDF-S catalog by \citet{Xue:2016}, where we use $\Lx=0.721 L_{[0.5-7.0\rm{keV}]}$ \citep{Xue:2016}.
The majority of our sources are located at $0.7<z<1.7$ where H$\alpha$ is within the spectral range. Both H$\beta$ and \ion{Mg}{2} are significantly weaker and thus more challenging to detect. Within the redshift range $1.2<z<1.7$ where we are able to simultaneously cover both H$\alpha$ and H$\beta$ we only detect H$\beta$ reliably in $\sim26$\% (35/135) of the cases. 

For four of our AGN we found previous near-IR spectroscopic observations in the literature. CID~87 (XID~18) and LID~1646 (XID~5321) have been observed with VLT/XSHOOTER \citep{Bongiorno:2014,Brusa:2015}, as a result of a selection based on their red optical colors ($R-K>4.5$). While the S/N and spectral resolution of the XSHOOTER spectra is higher, our results for these two objects are consistent with their work. We note however that our measurement for CID~87 has large uncertainties. 

For CID~352, our FMOS spectrum covers H$\beta$, while \citet{Trakhtenbrot:2016} observed the H$\alpha$ line region in $K$-band with MOSFIRE at Keck. Finally, 
for CID~113 at $z=3.3$ we detect the \ion{Mg}{2} line in $J$-band, while \citet{Trakhtenbrot:2016} presents MOSFIRE observations of the H$\beta$ line in $K$-band. We note that our SMBH mass estimates are higher by 0.5 and 0.3~dex  than those reported in their work for CID~352 and CID~113 respectively. For CID~352 this difference is mainly due to the use of a different virial mass formula, while for CID~113 it is caused by a broader \ion{Mg}{2} line in our observations compared to the expectation based on the H$\beta$ line in \citet{Trakhtenbrot:2016}.
 Two of the COSMOS AGN in our sample at $z>2$ (CID~166 and CID~346) are also part of the ESO Large Programme SUPER (PI: V. Mainieri, Circosta et al., in prep). SUPER will provide high S/N and spatially resolved observations in $H$ and $K$-band using VLT/SINFONI.

\begin{figure}
\centering
\resizebox{\hsize}{!}{ \includegraphics[clip]{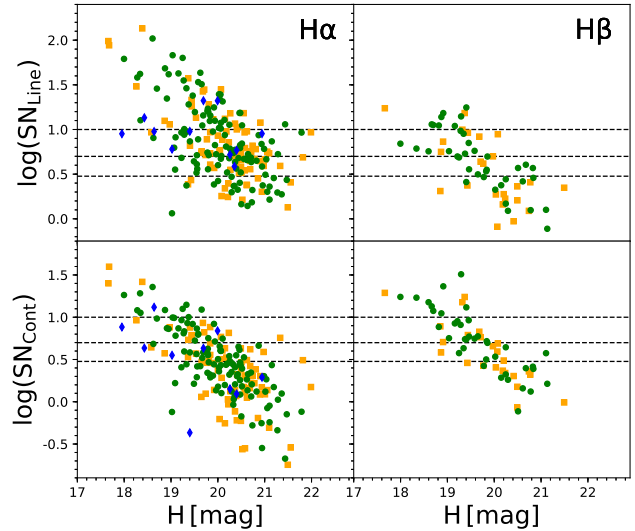} }
\caption{Signal to noise (S/N) ratios for our sample as a function of $H$ magnitude for the H$\alpha$ (left panels ) and H$\beta$ (right panels) region. The upper panels give the S/N at the peak of the respective emission line and the lower panels show the continuum S/N per 5~\AA{} wavelength pixel.  We mark the three surveys: COSMOS (green circles), SXDS (orange squares) and E-CDF-S (blue diamonds). 
The horizontal dashed lines indicate S/N ratios of 3, 5 and 10.}
\label{fig:sn} 
\end{figure}

In Figure~\ref{fig:sn} we show the S/N ratios of our sample, for each emission line and the continuum, as a function of $H$-band magnitude. We define the line S/N (S/N$_{\rm{Line}}$) as the signal to noise per pixel measured at the peak of the respective broad emission line, while the continuum S/N (S/N$_{\rm{Cont}}$) is given by the median S/N per wavelength pixel over two 40\AA{} wide regions red-ward and blue-ward of the respective emission line (16 resolution elements). For the HR spectra, we re-binned the spectra to the same resolution as the LR spectra before computing the S/N. The majority of our sample has S/N$_{\rm{Cont}}<10$, some of them have S/N$_{\rm{Cont}}<1$. At least for the broad H$\alpha$ line S/N$_{\rm{Line}}>3$ for the majority of our objects, allowing reliable measurements in most cases. Measurements for the considerably weaker H$\beta$ line are overall much less reliable and should be taken with caution, especially for fainter objects ($H>20$~mag). Thus, we emphasize that the large size of our sample comes at the cost of typically larger uncertainties for individual objects. We investigate and discuss the typical uncertainty of our measurements in sections~\ref{sec:LRHR} and \ref{sec:linecomp}.

\begin{figure*}
\centering
\includegraphics[width=8.8cm,clip]{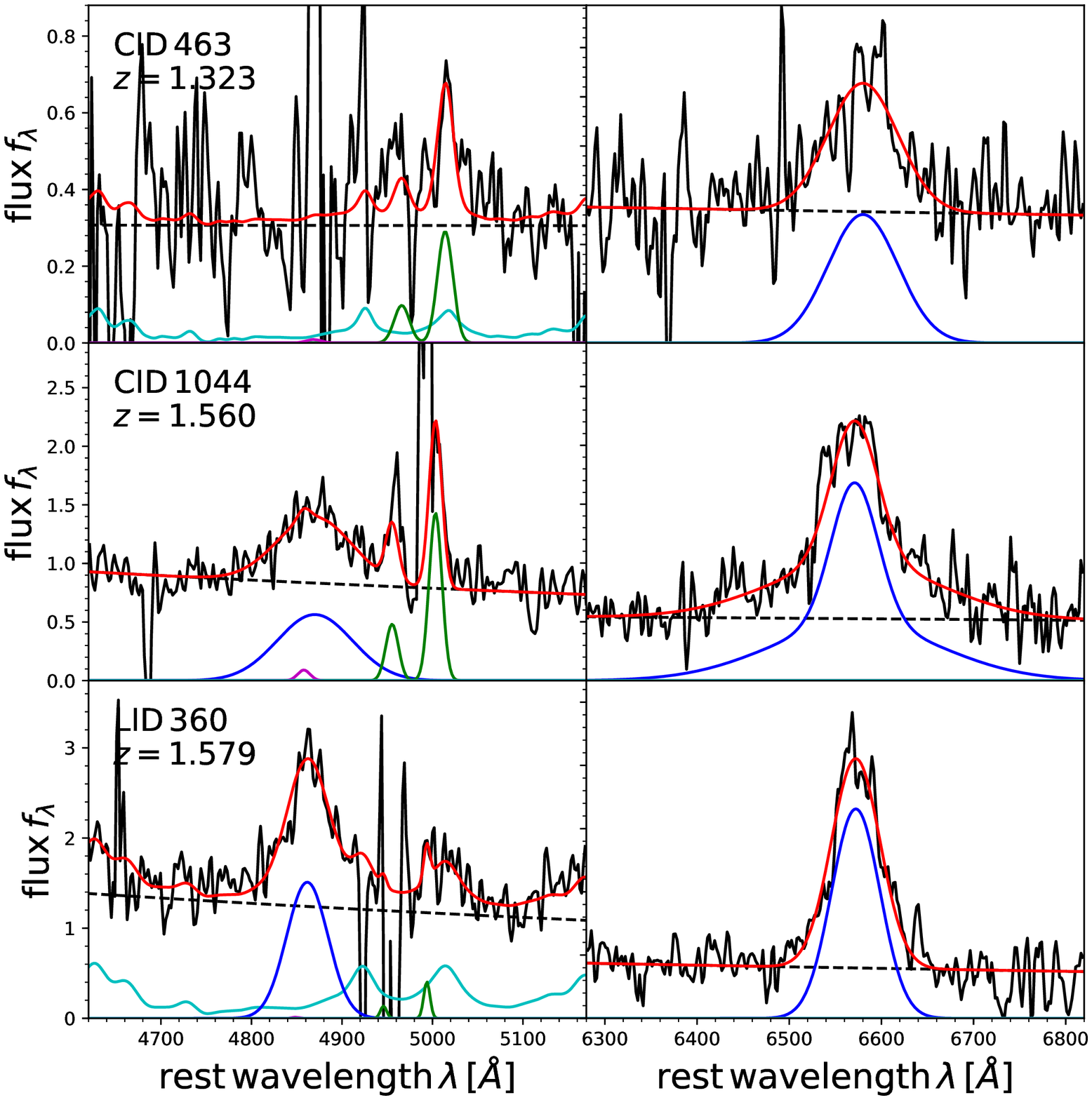} 
\includegraphics[width=8.8cm,clip]{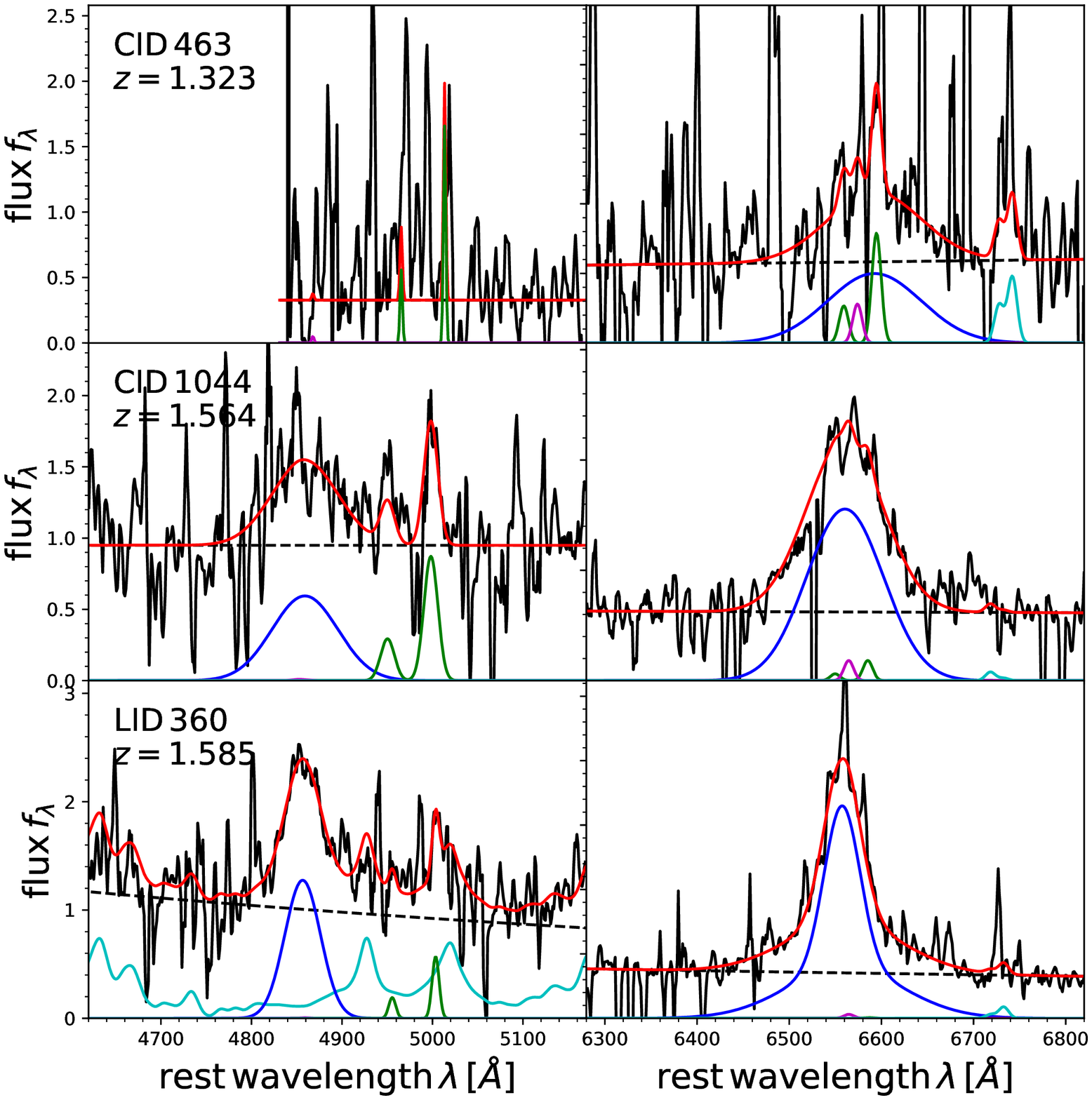}
\caption{Example rest-frame FMOS spectra and best-fit spectral model for COSMOS AGN. The left panels show spectra obtained in LR mode, while the right panels show HR mode spectra for the same objects. The red line indicates the total best-fit spectral model, the black dashed line represents the power law continuum, the blue line is for the broad Balmer lines, the narrow Balmer lines are shown in magenta. In the H$\beta$ fit (left sub-panels) green lines indicate [\ion{O}{3}] and cyan shows the broadened iron template. In the H$\alpha$ (right sub-panels) green lines indicate [\ion{N}{2}] and cyan shows [\ion{S}{2}]. For CID~463 no broad H$\beta$ has been detected.}
\label{fig:spec_cos} 
\end{figure*}

\begin{figure*}
\centering
\includegraphics[width=8.8cm,clip]{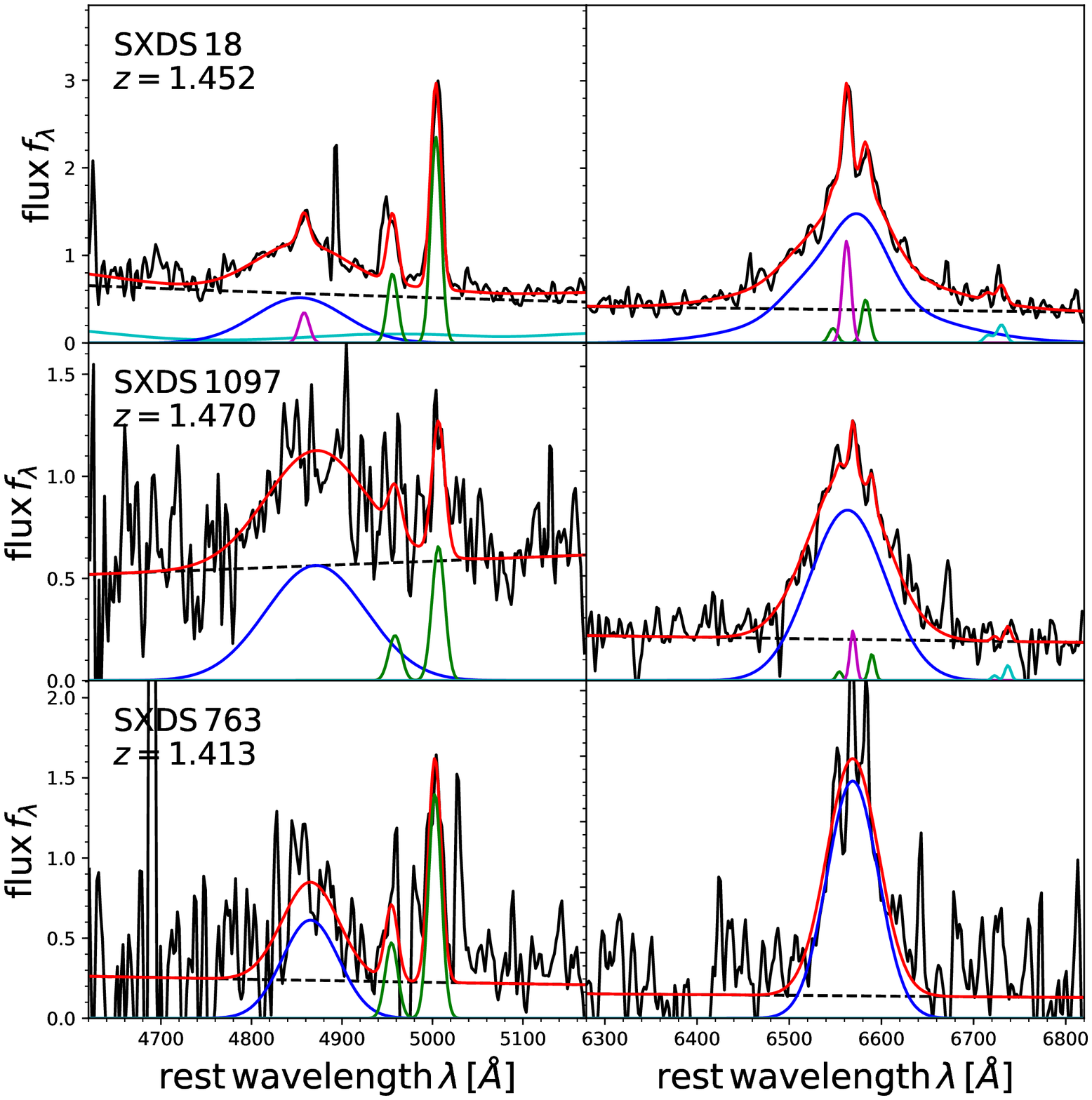} 
\includegraphics[width=8.8cm,clip]{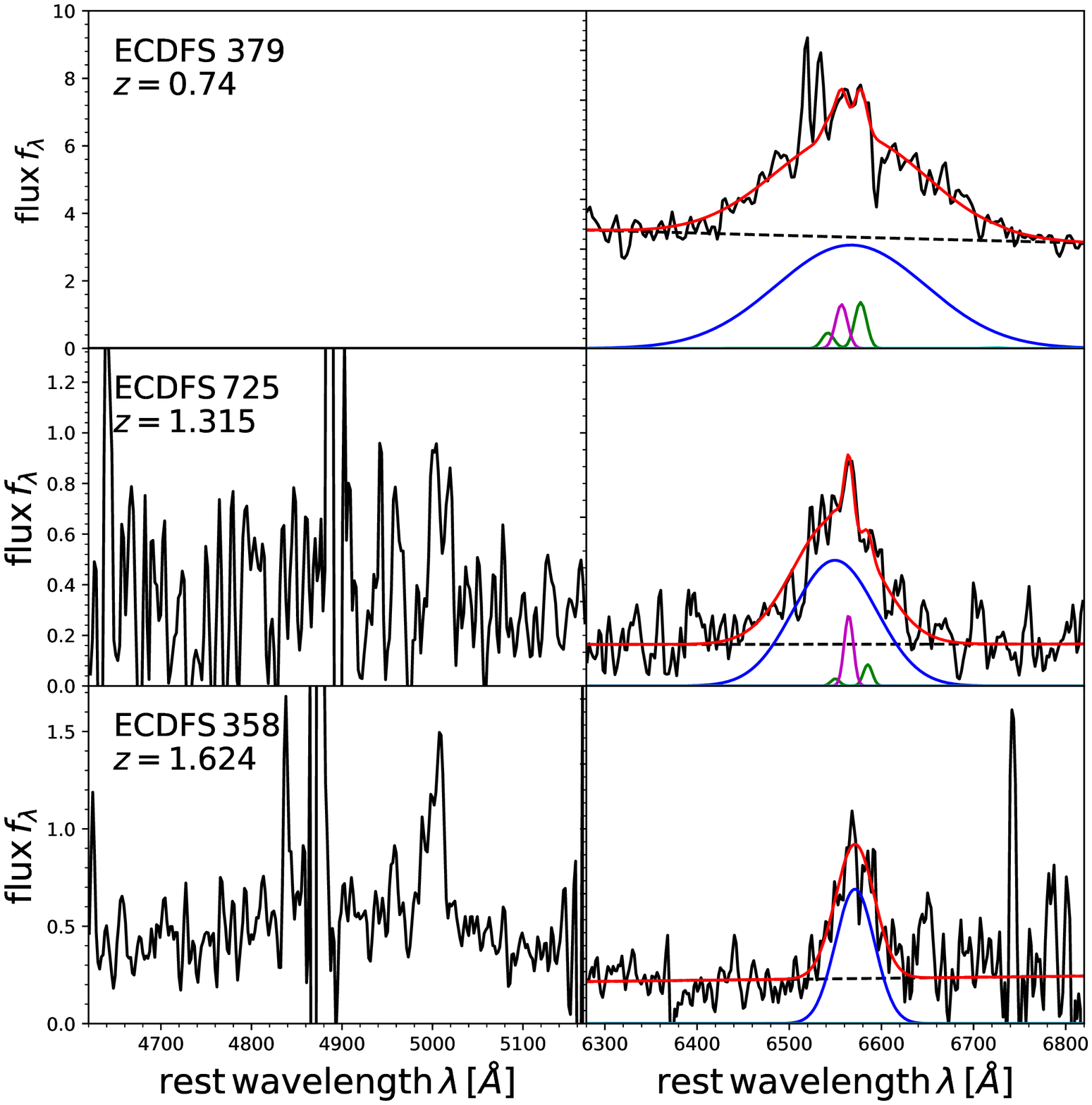}
\caption{Same as Figure~\ref{fig:spec_cos} for example spectra in SXDS (left panels) and E-CDF-S (right panels). In E-CDF-S we detected H$\beta$ in none of the spectra.}
\label{fig:spec_sxds} 
\end{figure*}


\begin{figure}
\centering
\resizebox{\hsize}{!}{ \includegraphics[clip]{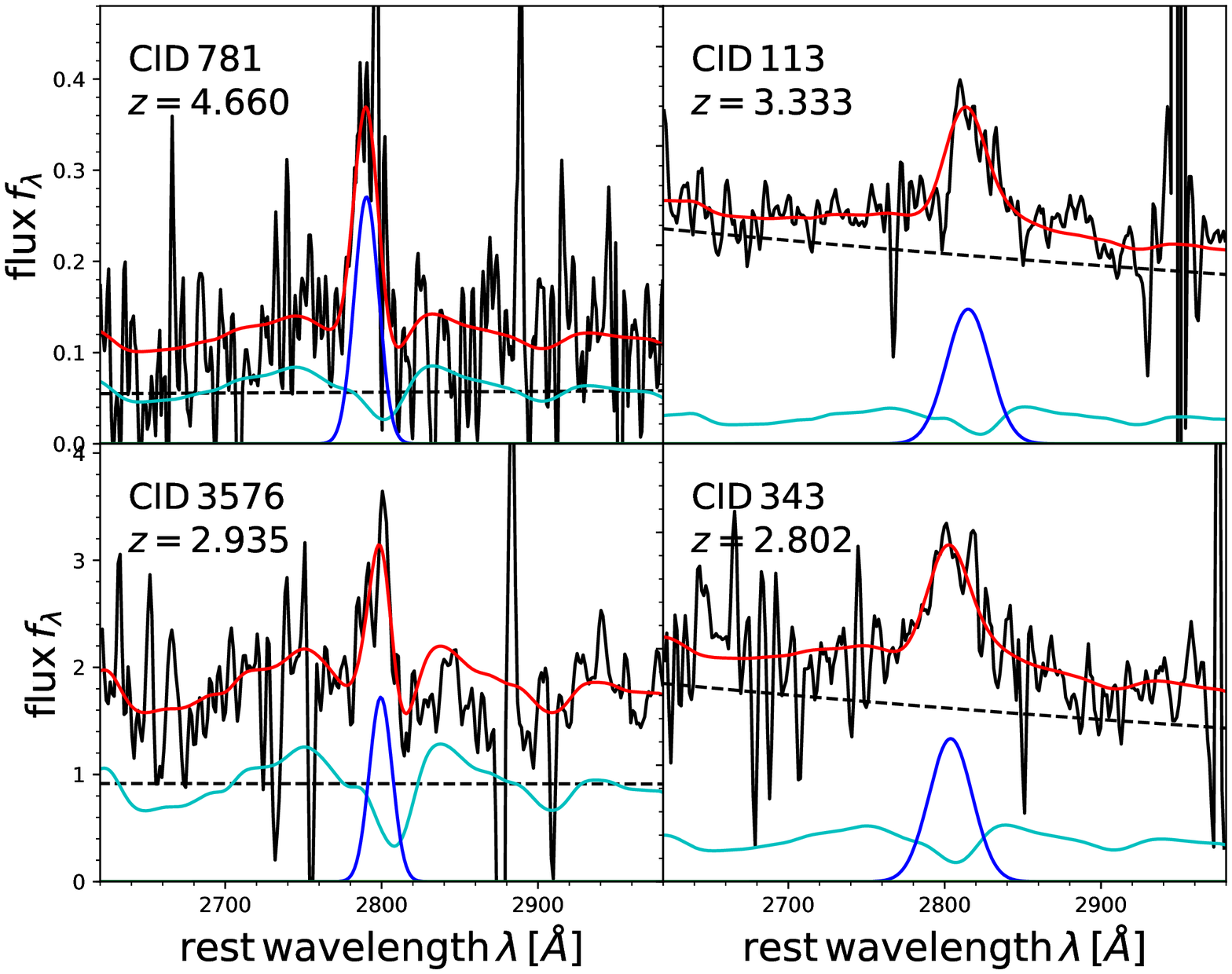} }
\caption{Spectra covering the \ion{Mg}{2} line region for the four AGN in COSMOS at $z>2.8$ where a broad \ion{Mg}{2} line is detected. We show the total best-fit model (red line) as well as  the individual components consisting 
of power law continuum (black dashed line), broad \ion{Mg}{2} line (blue line) and broadened iron template (cyan line).}
\label{fig:spec_mgii} 
\end{figure}

\section{Analysis}  \label{sec:analysis}
\subsection{Spectral measurements} \label{sec:line}
Our measurements of H$\alpha$, H$\beta$ and \ion{Mg}{2}, the continuum fluxes and the narrow [\ion{O}{3}] lines are based on spectral model fits to the respective wavelength regions.
Our procedure for continuum fitting and emission line modeling is similar to several previous studies \citep[e.g.][M13]{Schulze:2010,Shen:2011,Shen:2012b,Schulze:2017}. 

We correct the spectra for galactic extinction using the extinction map from \citet{Schlegel:1998} and the reddening curve from \citet{Cardelli:1989} and shift them to their rest frame using their spectroscopic redshift from either the X-ray catalog \citep{Marchesi:2016,Akiyama:2015,Hsu:2014} or the FMOS catalog \citep{Silverman:2015}. For the spectral fit, we mask out regions in the near-IR which are strongly affected by OH emission.
We initially fit every FMOS spectrum by both a narrow line and  a broad-line AGN model, based on a Levenberg-Marquardt least-squares minimization as implemented in MPFIT \citep{Markwardt:2009}. We classify the AGN as type-1 if the broad-line AGN model leads to a significant improvement in the reduced $\chi^2$, i.e. a decrease in reduced $\chi^2$ by at least 25\%. Otherwise we classify the object as a type-2 AGN. We visually inspect each spectral fit and manually adjust the model fit if appropriate. 
For the purpose of this paper, we only further consider those AGN for which our analysis of the FMOS spectrum results in a type-1 classification. \rev{For AGN with FWHM$<2000$~km/s there exist the possibility of confusion with an intrinsically type-2 AGN or with an AGN-driven wind \citep{ForsterSchreiber:2018}, especially for the LR mode data. However, the inclusion of ancillary data from optical spectroscopy and the SED minimizes this possibility.}

We here provide details on the spectral models for the individual line complexes \citep[see also][]{Schulze:2017}.
For H$\alpha$, we first fit a local power-law continuum to wavelength regions free from emission lines. An emission line model is fit to the continuum subtracted spectrum within $6200-7000$\AA{}. In the case of the narrow line model only the narrow emission lines are included, composed of single Gaussians each for H$\alpha$, [\ion{N}{2}] $\lambda,\lambda6548,6584$ and [\ion{S}{2}] $\lambda,\lambda6717,6731$. The line widths and offsets of all narrow lines are constrained to the same value (in velocity space). Their flux ratios are free to vary, apart from the flux ratio of the [\ion{N}{2}] lines, which is fixed to 2.96. For the broad-line model in addition the broad H$\alpha$ line is fit by up to three Gaussians, which is flexible enough to capture the often non-Gaussian broad-line profile. \rev{If a broad line model is preferred, we test both a model with only a broad line and one with the inclusion of narrow lines. We use the latter model if it leads to a decrease of the reduced $\chi^2$ by at least 25\%, and a broad line only model otherwise.}

For H$\beta$, the narrow-line model is composed of a local power-law continuum, a single narrow Gaussian for H$\beta$ and a single Gaussian each to model the narrow  [\ion{O}{3}] $\lambda\lambda4959,5007$ lines with their line ratio  fixed to 3.0. The velocity offset of all narrow lines are tied together and their line widths are constrained to the same value.
For the broad-line H$\beta$ model we fit a local pseudo-continuum, consisting of a power-law continuum and an optical iron template \citep{Boroson:1992}, broadened by a Gaussian, whose width is a free parameter. A line model is fit to the pseudo-continuum subtracted spectrum over the range $4700-5100$\AA\,. We fit up to three Gaussians for the broad H$\beta$ line and include the narrow H$\beta$  and [\ion{O}{3}] lines as above.

We detect broad \ion{Mg}{2} in four cases in our FMOS sample. To model their spectrum, we again first fit and subtract a local pseudo-continuum, consisting of a power-law and iron contribution. For the latter we use a broadened \ion{Fe}{2} template from \citet{Mejia:2016}. We then fit the emission line model over the range $2700-2900$\AA{}. A single broad Gaussian is sufficient to model the line profile in all cases.

In section~\ref{sec:linecomp}, we include optical spectra covering the \ion{Mg}{2} line at $z<1.7$ from the optical spectra of SDSS and zCOSMOS \citep[bright and deep component;][]{Lilly:2007}.
The \ion{Mg}{2} line fitting procedure for these optical spectra in COSMOS is outlined in \citet{Schramm:2013} and M13. 
A detailed discussion of SMBH masses obtained from optical spectroscopy in COSMOS and their corresponding SMBH mass catalog will be presented in a future publication. For the SXDS sample the \ion{Mg}{2} measurements are taken from N12 and for the E-CDF-S sources from M13. These measurements are based on a similar spectral fitting method, which warrants to combine them in our analysis.


For every emission line, we use the best-fit model to measure the FWHM (corrected for the instrumental resolution) of the broad and/or narrow component, the line flux and the velocity offset. In addition, we measure the continuum flux from the power-law continuum fit and use these to compute the monochromatic continuum luminosities at rest frame 5100\AA{} and 3000\AA{}.
We derive uncertainties on these parameters based on 100 Monte-Carlo realizations for each spectrum \citep[e.g.][]{Shen:2011,Mejia:2016,Schulze:2017}. For each realization, we modify the spectrum by adding Gaussian random noise, with the standard deviation at each pixel drawn from the flux density error spectrum, and re-fit this modified spectrum. The uncertainties for each measured parameter are taken as the 68\% range from the distribution of this parameter measured on the best-fit to the set of mock spectra. This approach generally provides more realistic uncertainties than the formal errors of the fitting procedure.

We have re-measured the systemic redshifts of the AGN from the best-fit spectral models. In cases where our spectroscopy covers the [\ion{O}{3}] lines and [\ion{O}{3}]$\lambda5007$ is detected at S/N$>5$ we use the  model peak of that line as our systemic redshift estimate. Otherwise, we use the model peak of the total H$\alpha$ profile or if H$\alpha$ is not covered of the total H$\beta$ profile as systemic redshift. These lines have average offsets from the systemic redshift by less than $120$~km s$^{-1}$ \citep{Shen:2016b}.
Our near-IR redshift measurements are given in Table~\ref{tab:prop}, together with the catalog redshift which is largely based on optical spectroscopy  \citep{Marchesi:2016,Akiyama:2015,Xue:2011}. We generally find a good agreement between the catalog redshift and the near-IR redshift. We define the difference in the redshift measures as  $c(z_{\rm{cat}}-z_{\rm{NIR}})/(1+z_{\rm{NIR}})$. For the combined sample, find a median of 28 and median absolute deviation (MAD) of 228 km~s$^{-1}$.

The continuum and line measurements are given in Table~\ref{tab:prop}. Example line fits for all three surveys, as well as for COSMOS LR and HR spectra are shown in Figure~\ref{fig:spec_cos} and Figure~\ref{fig:spec_sxds}. The spectra and best-fits for the \ion{Mg}{2} line regions for the four objects detected are shown in Figure~\ref{fig:spec_mgii}.

We note that our new measurements for H$\alpha$ are consistent with those presented before in M13 and N12. Comparing with the sample in M13, we find for the difference in $\log$~FWHM and $\log L_{\rm{H}\alpha}$ a mean and standard deviation of ($0.000, 0.168$) and ($-0.012, 0.173$), respectively. Compared to the H$\alpha$ measurements in N12 we find a mean and standard deviation in $\log$~FWHM of ($0.011, 0.082$). N12 do not provide measurements of $L_{\rm{H}\alpha}$ but rather couple their H$\alpha$ measurements to the virial mass estimator for \ion{Mg}{2}.

\begin{deluxetable*}{lccl}
\tabletypesize{\scriptsize}
\tablecaption{Spectral measurements for FMOS H$\alpha$ and H$\beta$ sample}
\tablewidth{18cm}
\tablehead{
\colhead{Column name} & \colhead{Format} & \colhead{Units} & \colhead{Description}
}
\startdata
Field & String & & Survey field \\
XID & String & & X-ray identifier in the format "survey\_id"  with survey in (CID,LID,SXDS,E-CDF-S)\\
XMM & Int &  & ID from XMM-COSMOS  for COSMOS objects, X-ray ID otherwise\\
RAdeg & Float & deg & Optical Right Ascension (J2000) \\
DEdeg & Float & deg & Optical Declination (J2000) \\
z & Float &  & catalog redshift \\
zsys & Float &  & redshift measured from peak of lines in near-IR spectra \\
f\_zsys & String &  & line used for near-IR readshift measurement \\
imag & Float & mag & i-band AB magnitude \\
Jmag & Float & mag & J-band 3" aperture AB magnitude \\
Hmag & Float & mag & H-band 3" aperture AB magnitude \\
E(B-V) & Float & mag & Galactic extinction \\
logLx & Float & erg/s & 2-10 keV rest frame luminosity \\
E(B-V)i & Float & & estimated $E(B-V)$ due to intrinsic dust reddening \\
HaMode & String &  & FMOS mode for H$\alpha$ \\
HaSN & Float &  & median continuum S/N per pixel around H$\alpha$ \\
HaLSN & Float &  & peak S/N of the H$\alpha$ line \\
logLHa & Float & erg/s & broad H$\alpha$ line luminosity \\
logLHadr & Float & erg/s & reddening corrected broad H$\alpha$ line luminosity \\
e\_logLHa\ & Float & erg/s & measurement error for broad H$\alpha$ line luminosity \\
FWHMHa & Float & km/s & broad H$\alpha$ FWHM \\
e\_FWHMHa & Float & km/s & measurement error for broad H$\alpha$ FWHM \\
logMHa & Float & $M_\odot$ & SMBH mass estimated from H$\alpha$ \\
e\_logMHa & Float & $M_\odot$ & measurement error for SMBH mass estimated from H$\alpha$ \\
logLbolHa & Float & erg/s & bolometric luminosity estimated from H$\alpha$ \\
e\_logLbolHa & Float & erg/s & measurement error for bolometric luminosity estimated from H$\alpha$ \\
logEddRHa & Float &  & Eddington ratio estimated from H$\alpha$ \\
e\_logEddRHa & Float &  & measurement error for Eddington ratio estimated from H$\alpha$ \\
HbMode & String &  & FMOS mode for H$\beta$ \\
HbSN & Float &  & median continuum S/N per pixel around H$\beta$ \\
HbLSN & Float &  & peak S/N of the H$\beta$ line \\
logLO3 & Float & erg/s & [OIII] line luminosity  \\
e\_logLO3 & Float & erg/s & measurement error for [OIII] line luminosity  \\
logLHb & Float & erg/s & broad H$\beta$ line luminosity \\
logLHbdr & Float & erg/s & reddening corrected broad H$\beta$ line luminosity \\
e\_logHb & Float & erg/s & measurement error for broad H$\beta$ line luminosity \\
logL5100 & Float & erg/s & continuum luminosity at 5100 Angstroem \\
logL5100dr & Float & erg/s & reddening corrected continuum luminosity at 5100 Angstroem \\
logL5100cr & Float & erg/s & continuum luminosity at 5100 Angstroem with average host correction and reddening correction\\
e\_logL5100 & Float & erg/s & measurement error for continuum luminosity at 5100 Angstroem \\
FWHMHb & Float & km/s & broad H$\beta$ FWHM \\
e\_FWHMHb & Float & km/s & measurement error for broad H$\beta$ FWHM \\
logMHb & Float & $M_\odot$ & SMBH mass estimated from H$\beta$ \\
e\_logMHb & Float & $M_\odot$ & measurement error for SMBH mass estimated from H$\beta$ \\
logLbolHb & Float & erg/s & bolometric luminosity estimated from H$\beta$ \\
e\_logLbolHb & Float & erg/s & measurement error for bolometric luminosity estimated from H$\beta$ \\
logEddRHb & Float &  & Eddington ratio estimated from H$\beta$ \\
e\_logEddRHb & Float &  & measurement error for Eddington ratio estimated from H$\beta$ \\
\enddata
\tablecomments{Format for the sample properties and spectral measurements for the FMOS sample with either broad H$\alpha$ or H$\beta$ detection. This table is available in its entirety in a machine-readable form in the online journal. A portion is shown here for guidance regarding its form and content. In addition, we make it available online under the following link: \href{http://member.ipmu.jp/fmos-cosmos/fmosBL.fits}{http://member.ipmu.jp/fmos-cosmos/fmosBL.fits}. }
\label{tab:prop}
\end{deluxetable*}

\subsection{Spectral energy distribution and reddening} \label{sec:sed}
To investigate the amount of dust reddening affecting our AGN sample, we construct the spectral energy distribution (SED) for the AGN sample in COSMOS and SXDS. For COSMOS, we utilize the multi-wavelength catalog COSMOS2015 \citep{Laigle:2016}, where we include homogenized photometry in $u^*$, $B$, $V$, $g$, $r$, $i$, $z+$ from the Canada-France-Hawaii Telescope (CFHT) and Subaru Suprime-Cam, in $YJHK_s$ from VIRCAM/VISTA (UltraVISTA DR2) and in $3.6\mu m$, $4.5\mu m$, $5.8\mu m$, and $8.0\mu m$ from {\it Spitzer}-IRAC  (S-COSMOS and SPLASH). For SXDS we use the homogenized multi-wavelength catalog by \citet{Mehta:2018}. We include data only from the CFHT $u$-band program MUSUBI, Hyper Suprime-Cam Subaru Strategic Program (HSC-SSP; $grizY$), the VISTA Deep Extragalactic Observations (VIDEO) survey ($YJHK_s$) and {\it Spitzer}-IRAC. We use photometry within a 3\arcsec aperture.

\begin{figure}
\centering
\resizebox{\hsize}{!}{\includegraphics[width=14cm,clip]{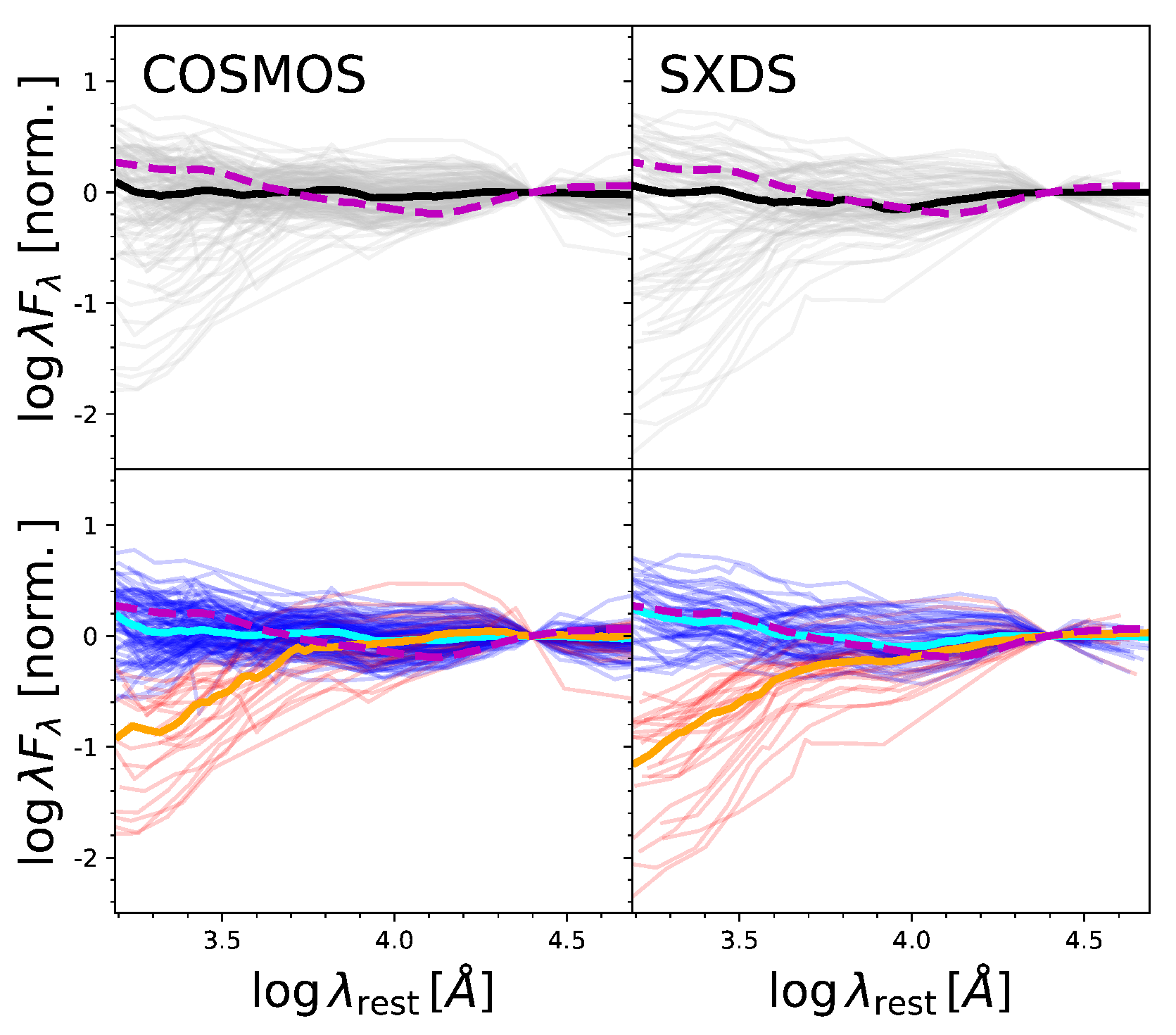} }
\caption{Spectral energy distribution (SED) for the FMOS sample in COSMOS (left) and SXDS (right), normalized at 2.5$\mu$m. The gray lines in the upper panels show the SED for individual objects, while the median SED is shown as black solid line. The type-1 AGN SED template from \citet{Richards:2006} is shown for comparison (magenta dashed line). In the lower panels we split the FMOS sample into normal blue AGN (blue lines; cyan line for median) and red AGN (red lines; orange line for median), based on their SED shape as described in the text.}
\label{fig:sed} 
\end{figure}

We show the SEDs for the individual broad-line AGN in the FMOS sample in COSMOS and SXDS in the upper panels of Figure~\ref{fig:sed}. We normalize each SED to $\lambda F_\lambda = 1$ at $2.5\mu$m. 
Although the host galaxy can make a significant contribution at $\sim1\mu$m, the emission at this wavelength is expected to be dominated by the AGN emission and to be unaffected by dust reddening. In addition, we plot the median SED for the FMOS  sample (black solid line) and the unobscured quasar SED template from \citet{Richards:2006}, based on SDSS quasars with $45<\log L<46.02$ (the fainter half of their sample; magenta dashed line). This SED template overall agrees well with the typical SEDs of our AGN, while some differences are present. Comparing our median SED to the \citet{Richards:2006} quasar SED, there is enhanced flux around $1-2\mu$m, which we attribute to host galaxy emission \citep[see e.g.][]{Bongiorno:2012}. Furthermore, at shorter wavelengths the spectral slope is redder. There is a significant population with much redder slopes than both the median SED of the FMOS sample and the \citet{Richards:2006} SED. A plausible explanation for this population is dust reddening within the host galaxy. 

We define the reddened AGN population in the FMOS sample as deviating by more than 0.5~dex in their rest-frame UV to near-IR flux ratio from the \citet{Richards:2006} quasar SED, the typical dispersion observed in  \citet{Richards:2006}. Specifically, we use $\log (\lambda F_{\lambda}(2500{\rm \AA{}})/ \lambda F_{\lambda}(2.5\mu{\rm m}) )<-0.3$. We indicate this separation into normal blue AGN and red AGN in the lower panels of Figure~\ref{fig:sed} by blue and red color respectively. In SXDS 27/87 AGN are classified as red, while in COSMOS we identify 27/141 AGN to have a red SED. In the small E-CDF-S sample, none of the objects are identified as a red AGN.
The blue AGN population is consistent with a standard un-reddened AGN SED within their typical dispersion, while the red AGN population is consistent with dust reddening by $E(B-V)\gtrsim0.1$. We estimate the amount of reddening for each object by de-reddening their observed SEDs with an SMC-like dust extinction curve \citep{Gordon:2003}\footnote{\rev{We note that also a stepper extinction curve has been suggested for a few luminous red QSOs \citep{Zafar:2015}. Our results do not significantly depend on the specific choice of extinction law, thus we here adopt the most commonly used SMC extinction curve. A more detailed study of the extinction curve for AGN is beyond the scope of this paper.}} to match the median SED of the blue AGN population in the respective field, using  $\chi^2$ minimization. We derive dust extinction values in the range $E(B-V)=0.1-0.65$, which are listed in Table~\ref{tab:prop}. 

For the red AGN population, we correct the continuum and broad emission line luminosities obtained from the spectral fits for dust reddening using the obtained $E(B-V)$ estimates. We verified that correcting the entire spectrum of each of the red AGN does not affect the best-fit spectral parameters in a significant way.
For H$\alpha$, the reddening correction has a moderate effect ($<0.5$~dex) on the line luminosity, but it can be up to more than 1~dex for $L_{3000}$. We note that the default black hole mass estimates and bolometric luminosities for the majority of our sample are based on H$\alpha$. For those AGN where we measure a broad H$\beta$ line, only 3 are classified as red AGN. For these 3 objects we do not apply a reddening correction for the narrow [\ion{O}{3}] emission lines.

\begin{figure*}
\centering
\includegraphics[width=18cm,clip]{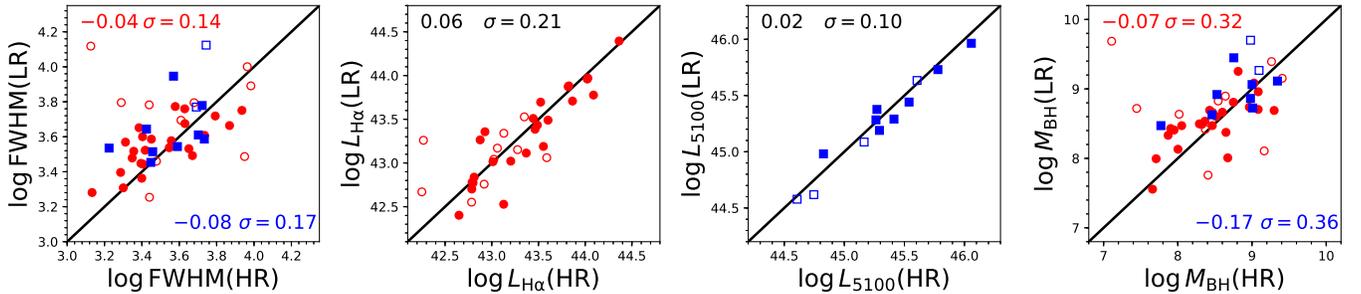} 
\caption{Comparison between FWHM, $L_{\rm{H}\alpha}$, $L_{5100}$ and \mbh\ measurements for the FMOS high resolution (HR) and low resolution (LR) spectra for AGN in COSMOS which have been observed in both modes. The red circles are for H$\alpha$ measurements and the blue squares are for measurements in the H$\beta$ region. Open symbols indicate cases in which at least for one of the two spectra the best-fit model suffers from low quality or large uncertainties. The black solid line is the one-to-one relation. We show the mean and standard deviation around the one-to-one relation, excluding the poor quality cases, in the corner of each panel.}
\label{fig:compline} 
\end{figure*}

\section{Results} \label{sec:results}
\subsection{Comparison of FMOS low vs. high resolution mode}  \label{sec:LRHR}
For the COSMOS sample, we have obtained spectroscopy in both the LR and HR mode of FMOS. While we will combine the samples from both campaigns further below, we carried out the spectral fitting independently on both sub-samples. There are in total 39 AGN in our COSMOS sample with observations in both modes. We are thus able to compare the emission line and continuum measurements from these two independent observations with their different spectral resolutions. This provides an independent assessment of the measurement uncertainty of broad-line widths and luminosities.

In Figure~\ref{fig:compline}, we show the comparison of H$\alpha$ and H$\beta$ FWHM, H$\alpha$ luminosity $L_{\rm{H}\alpha}$, continuum luminosity at 5100\AA{} $L_{5100}$ and the derived SMBH masses \mbh obtained from the LR and HR spectra. For the \mbh estimates we use the relations as defined below in section \ref{sec:mbh}, which are based on the former quantities. We indicate objects whose spectral fit is affected by poor data quality in one of the two spectra by open symbols. 
We see that the poor quality cases  tend to be outliers in their broad-line FWHM and to a lesser extend also in  $L_{\rm{H}\alpha}$. For the higher quality cases, we find a generally good correlation between the LR and HR results. The continuum luminosity $L_{5100}$ naturally shows the best agreement, while the line fits have larger dispersion. For the FWHM we find a standard deviation of 0.14~dex and 0.17~dex for H$\alpha$ and H$\beta$, respectively. In particular for H$\beta$ the FWHM from the LR mode tends to be overestimated compared to the HR spectra, with a mean offset of $-0.08$~dex. These trends propagate to the SMBH mass estimates, which show an uncertainty of $\sim0.35$~dex

The scatter in the measured line properties originates in ambiguities in the line fit for low S/N data, including the contribution of a narrow component or the choice of parametric model for the broad component, e.g. number of Gaussian components \citep{Denney:2009, Shen:2011,Karouzos:2015,Denney:2016}. We note that for the vast majority of our sources the continuum S/N for at least one of the spectra is $<5$. The dispersion we see in our sample is consistent with the expectation from previous studies for data of those S/N \citep{Denney:2009, Shen:2011}, which demonstrated that both the measurement scatter and systematic biases in the measurement will typically increase for spectra with continuum S/N$\lesssim5$. Another inherent source of scatter for these moderate-luminosity AGN is spectral variability \citep{Woo:2007,Denney:2009}.
\rev{The work by \citet{Sanchez:2017} studied the near-IR variability in the COSMOS field, finding $\sim32$\% of their broad-line AGN to show variability in $J$-band. For these, they report a structure function with a mean magnitude difference on a 1~yr timescale of $A=0.13$ and power-law index $\gamma=0.62$. For a typical time between the LR and HR observations of $2-5$~yr, this corresponds to an average variability of $0.08-0.14$~dex in luminosity and $0.02-0.03$~dex in FWHM. Thus, spectral variability is sub-dominant compared to the scatter due to data quality.}

We conclude that we expect an average uncertainty on the black hole mass estimates from our sample of $\sim0.3-0.35$~dex, due to data quality. This uncertainty adds in quadrature to the systemic uncertainty of $\sim0.3$~dex for the virial method itself.

We combine the measurements from the LR and HR sub-samples to build a merged COSMOS sample. In case both LR and HR spectra are available, we choose the best case, based on the S/N of the spectrum and visual inspection of the two best-fit models.

\subsection{Bolometric luminosities, Black hole masses and Eddington ratios} \label{sec:mbh}
We estimate black hole masses for our sample based on the virial method for broad-line AGN \citep[e.g.][]{McLure:2004, Greene:2005, Vestergaard:2006}. 
While the virial method is only calibrated based on reverberation mapping campaigns for the broad H$\beta$ line \citep{Vestergaard:2006,Collin:2006},  the broad \ion{Mg}{2} and  H$\alpha$ lines are known to provide reliable black hole mass estimates, when calibrated using AGNs having H$\beta$ based black hole mass estimates \citep{Greene:2005,Trakhtenbrot:2012,Shen:2012b,Mejia:2016}. Broad \ion{Mg}{2} has the advantage that it can be observed in optical spectra out to higher redshift ($z\sim2.3$) than H$\beta$ and with near-IR spectroscopy out to $z\sim6$ \citep[e.g.][]{Kurk:2007,Willott:2010,DeRosa:2011,Mazzucchelli:2017}. For our sample, the use of  \ion{Mg}{2} enables us to obtain black hole mass estimates for four AGN at $z>2.8$, including our highest redshift source CID~781 at $z=4.64$.

A major advantage of broad H$\alpha$ compared to H$\beta$ is that it is considerably stronger, which makes it a powerful alternative for low-luminosity AGN \citep{Greene:2007a,Dong:2012,Reines:2013}, but also for low S/N spectra, usually the case for near-IR spectroscopy. The latter is also realized for our FMOS sample, where the H$\beta$ line is generally detected at low to moderate quality, making H$\alpha$ the preferred \mbh estimator for our sample. Furthermore, the H$\alpha$ luminosity $L_{\rm{H}\alpha}$ is free of host galaxy contamination, contrary to the continuum luminosity $L_{5100}$. 
For some cases, an explicit comparison of both virial estimators is presented in section~\ref{sec:linecomp}.

We use H$\beta$ to anchor our virial mass estimators, i.e. ensure that the virial mass estimators for the other broad-lines are consistently calibrated to this H$\beta$ relation. We base our virial mass estimation on the relation for H$\beta$ by  \citet{Vestergaard:2006}, being directly calibrated to reverberation mapping studies
\begin{equation}
\mbh (\rm{H}\beta)= 10^{6.91} \left( \frac{L_{5100}}{10^{44}\,\mathrm{erg\,s}^{-1}}\right)^{0.5} \left( \frac{\mathrm{FWHM}}{1000\,\mathrm{km\,s}^{-1} }\right)^2 M_\odot  \label{eq:mbhHb}
\end{equation} 

At the moderate luminosities of our AGN sample, the host galaxy contamination to the continuum luminosity $L_{5100}$ is not negligible. \citet{Shen:2011} showed that host galaxy contamination becomes significant at $\log L_{5100}<45$ erg s$^{-1}$. We account for the host contribution in an average sense by applying the formula for the average host contamination given by \citet{Shen:2011} in their Equation~(1).
 
For H$\alpha$, we use the formula presented in \citet{Schulze:2017}
\begin{equation}
\mbh (\rm{H}\alpha)= 10^{6.71} \left( \frac{L_{\rm{H}\alpha}}{10^{42}\,\mathrm{erg\,s}^{-1}}\right)^{0.48} \left( \frac{\mathrm{FWHM}}{1000\,\mathrm{km\,s}^{-1} }\right)^{2.12} M_\odot.  \label{eq:mbhHa}
\end{equation} 

This black hole mass relation is based on Equation~\ref{eq:mbhHb}, and uses empirical scaling relationships between H$\alpha$ and H$\beta$ FWHM, as well as  between $L_{\rm{H}\alpha}$ and $L_{5100}$ from \citet{Jun:2015}. Their work updates the commonly used relations from \citet{Greene:2005} and extends them over a wider luminosity range.

For \ion{Mg}{2}, we use the relation from \citet{Shen:2011}, which is also tied to the virial estimator from \citet{Vestergaard:2006} \citep[see also][]{Trakhtenbrot:2012}
\begin{equation}
\mbh (\rm{MgII})= 10^{6.74} \left( \frac{L_{3000}}{10^{44}\,\mathrm{erg\,s}^{-1}}\right)^{0.62} \left( \frac{\mathrm{FWHM}}{1000\,\mathrm{km\,s}^{-1} }\right)^2 M_\odot
\end{equation} 


\begin{figure*}
\centering
\includegraphics[width=18cm,clip]{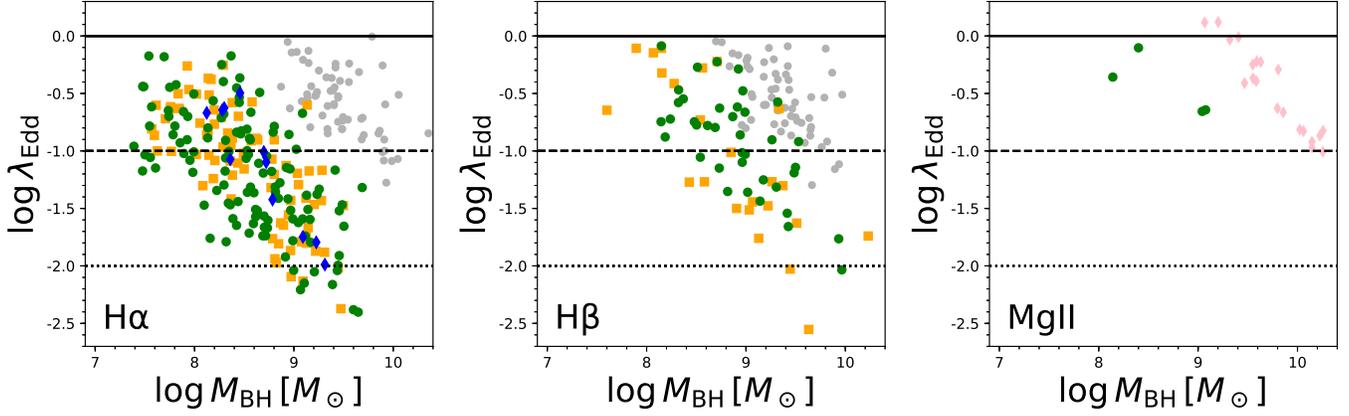} 
\caption{Distribution of the FMOS sample in the black hole mass - Eddington ratio plane for \mbh estimates based on  H$\alpha$ (left panel), H$\beta$ (middle panel) and \ion{Mg}{2} (right panel). The green circles are for COSMOS, the orange squares for SXDS and the blue diamonds for E-CDF-S. In addition, we show results for two luminous quasar samples from the literature. In the left and middle panels, the gray circles represent near-IR results on SDSS quasars from \citet{Shen:2012b}, and in the right panel we show as purple triangles \ion{Mg}{2} based \mbh\ and \er\ for luminous quasars at $z\sim3.5$ from \citet{Zuo:2015}. The horizontal black solid, dashed and dotted lines mark Eddington ratios of 1, 0.1 and 0.01 respectively.}
\label{fig:me} 
\end{figure*}


We base our estimate of the bolometric luminosity either on the continuum luminosities $L_{5100}$ and $L_{3000}$, or on the luminosity of the broad H$\alpha$ line $L_{\rm{H}\alpha}$. The intrinsic bolometric luminosity is given by the integration over the X-ray, UV and optical luminosity, i.e. by the contribution of the accretion disk and hot corona. The mid-IR emission should be excluded since it constitutes reprocessed UV-optical emission and would therefore lead to double-counting \citep[e.g.][]{Marconi:2004}. For $L_{5100}$, we use a constant bolometric correction factor $\rm{BC_{5100}}=7.0$ \citep{Netzer:2007b}, which is on average consistent with the luminosity-dependent bolometric correction of \citet{Marconi:2004} and excludes re-processed emission. For $L_{3000}$, we use the luminosity-dependent relation presented in \citet{Trakhtenbrot:2012}, inferred from the \citet{Marconi:2004} bolometric correction ($\rm{BC_{3000}}=4.12$ at $L_{3000}=10^{45}$ erg s$^{-1}$, the typical luminosity of our sample).
To obtain a bolometric luminosity estimate from broad H$\alpha$, we combine the scaling relation to $L_{5100}$ from \citet{Jun:2015} and $\rm{BC_{5100}}$, which gives
\begin{equation}
\log L_{\rm{bol}} = 0.96 (\log L_{\rm{H}\alpha} - 42) + 44.23   \ .  \label{eq:lbolHa}
\end{equation} 
We explicitly test the validity of these bolometric luminosity indicators in section~\ref{sec:lumcor}.

The Eddington ratio is given by $\er=L_{\rm{bol}}/L_{\rm{Edd}}$, where $L_{\rm{Edd}}\cong1.3\times 10^{38} (\mbh / M_\odot)$~erg s$^{-1}$ is the Eddington luminosity for the object, given its black hole mass.
We show the distribution of our sample in the \mbh-\er-plane in Figure~\ref{fig:me} for SMBH masses based on H$\alpha$, H$\beta$ and \ion{Mg}{2}. In total for our sample, 211 objects have robust black hole masses based on H$\alpha$, 63 based on H$\beta$ and 4 based on \ion{Mg}{2}. Particularly,  for H$\alpha$, we find a broad distribution in both black hole mass ($\log \mbh=[7.5,9.5]$) and Eddington ratio ($\log \er=[-2.5,0]$) with median values of $\log \mbh=8.54$ and $\log \er=-1.11$ and a dispersion of $0.54$~dex and  $0.51$~dex respectively. This is consistent with previous results on moderate-luminosity AGN in small area, deep fields \citep{Gavignaud:2008,Trump:2009,Merloni:2010,Schulze:2015,Suh:2015} and qualitatively agrees with the expectation from the underlying active black hole mass function (BHMF) and Eddington ratio distribution function \citep[ERDF; N12,][]{Schulze:2015}. Our H$\beta$ SMBH mass sample is shifted towards higher luminosities and therefore on average higher \mbh and \er. For this sample, we find median values of $\log \mbh=8.91$ and $\log \er=-0.90$. This is because the H$\beta$ mass sample extends to higher redshift $z>2$ and furthermore the detection of the weaker H$\beta$ line is only possible in the brighter subset of our sample at $z>1.2$. The few \ion{Mg}{2} detections are of AGN at even higher redshift $z>2.7$ and, given the common flux limit, preferentially target the more massive black holes at higher accretion rates than the  lower-$z$ samples, with median $\log \mbh=8.72$ and $\log \er=-0.50$.

We see the consequence of the flux limit on our sample by the apparent lack of objects in the lower left corner of the \mbh-\er-plane, at low \mbh and low \er. The  absence of AGN in our sample at high \mbh and high \er (upper right corner) is caused by the rarity of these objects \citep{Richards:2006, Kelly:2013,Schulze:2015}, which makes them effectively absent in the limited volume covered by COSMOS, SXDS and E-CDF-S. They will be found in large area surveys like SDSS \citep[e.g.][]{McLure:2004,Vestergaard:2009,Shen:2011}. We show in addition in Figure~\ref{fig:me} a sample of such luminous broad-line AGN at about the same redshift range with near-IR spectroscopy from \citet[][for H$\alpha$ and H$\beta$]{Shen:2012b} and \citet[][for \ion{Mg}{2}]{Zuo:2015}, targeted based on the SDSS quasar catalog \citep{Schneider:2010}. They indeed fill in the area in the upper-right corner not populated by our relatively small area survey. Combining such samples of luminous quasars with our sample of moderate-luminosity broad-line AGN provides a fairly complete coverage of the \mbh-\er-plane, which enables us to study the properties of a representative sample of type-1 AGN. The addition of the moderate-luminosity AGN regime, located  at the same redshift, is of special importance since it represents the bulk of the population.

\begin{figure*}
\centering
\includegraphics[width=12cm,clip]{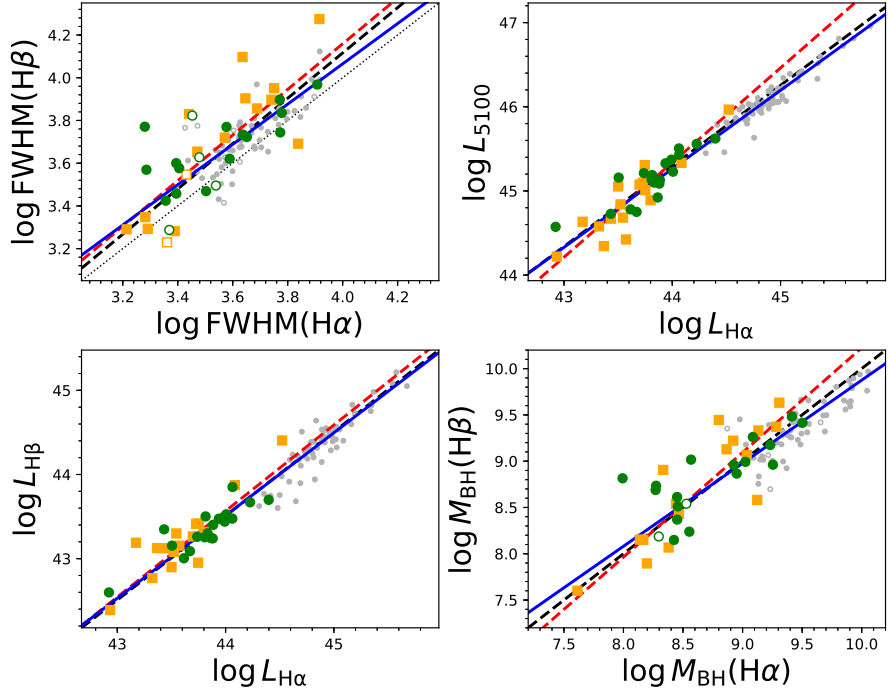} 
\caption{Comparison of  spectral measurements and derived black hole masses between H$\alpha$ and H$\beta$ for those AGN with both lines detected in COSMOS (green squares) and SXDS (orange circles), supplemented by the luminous quasar sample by \citet[][gray circles]{Shen:2012b}. The correlations for FWHM, $L_{\rm{H}\alpha}-L_{5100}$, $L_{\rm{H}\alpha}-L_{\rm{H}\beta}$ and \mbh\ are shown. In each panel, the black dashed line is for a reference relation as discussed in the text. The red dashed line is the best-fit to the FMOS sample and the blue solid line is the best-fit to the combined sample from FMOS and  \citet{Shen:2012b}. The black dotted line in the FWHM panel is the one-to-one relation.
}
\label{fig:compBalmer} 
\end{figure*}

\begin{figure*}
\centering
\includegraphics[width=18cm,clip]{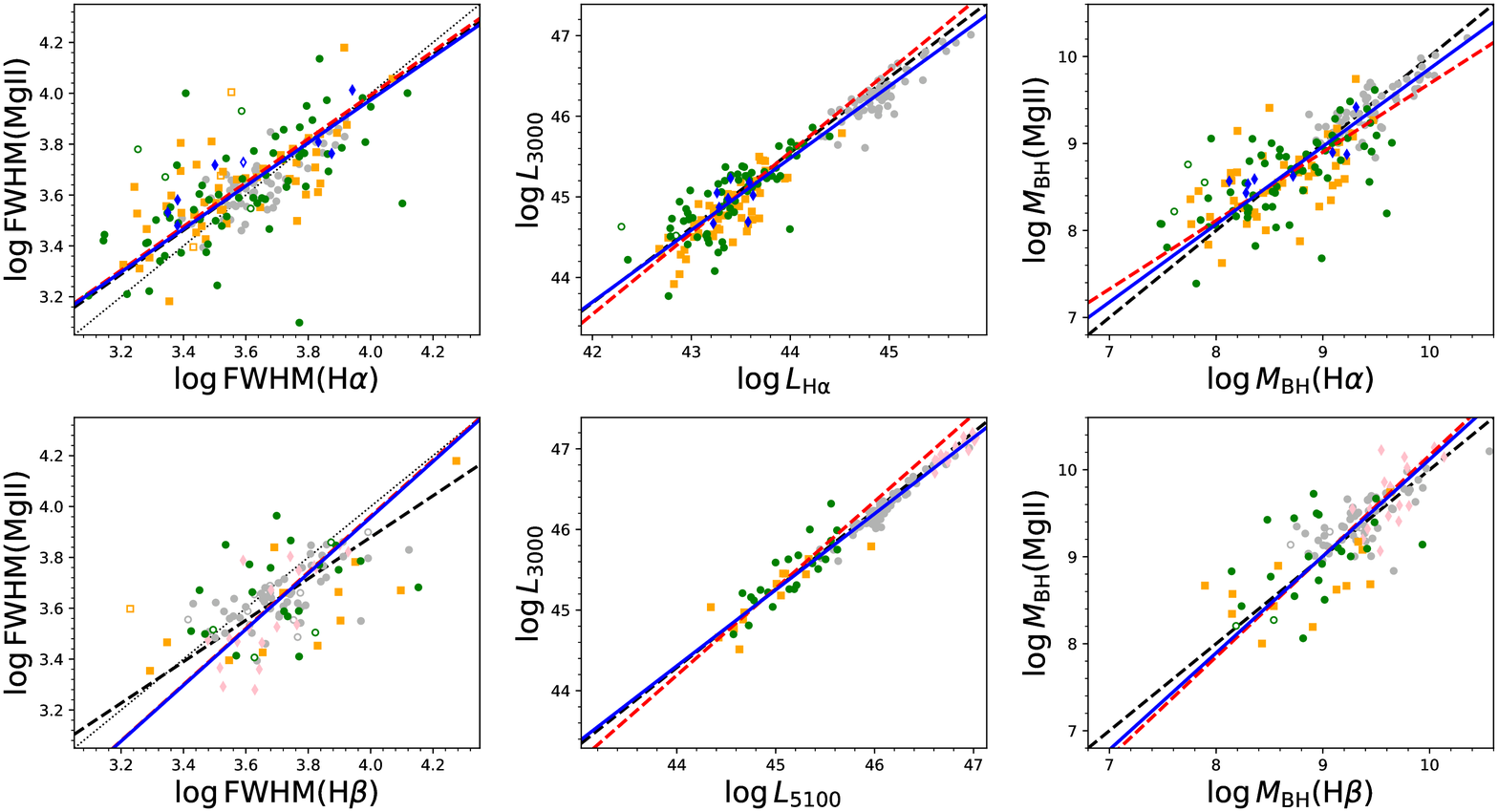} 
\caption{Comparison of  spectral measurements and derived black hole masses between H$\alpha$/H$\beta$  and \ion{Mg}{2} for those AGN with \ion{Mg}{2} measurements from optical spectra in COSMOS (green squares), SXDS (orange circles) and E-CDF-S (blue diamonds), augmented by the luminous quasar samples by \citet[][gray circles]{Shen:2012b} and \citet[][pink diamonds]{Zuo:2015}. The black, red and blue lines are for the reference relation, the best-fit to the FMOS sample and the best-fit to the combined sample of FMOS and the two luminous quasar samples. The black dotted line in the FWHM panel is the one-to-one relation.
}
\label{fig:compMgII} 
\end{figure*}

\subsection{Line correlations} \label{sec:linecomp}
In this section, we investigate the correlations of FWHM and luminosity and a comparison of virial black hole mass estimators. A significant amount of work has been put into the empirical establishment of these correlations and the cross-calibration of the virial method for different broad-lines \citep[][M13]{McGill:2008,Wang:2009,Assef:2011,Trakhtenbrot:2012,Shen:2012b,Ho:2012,Park:2013,Jun:2015,Mejia:2016,Woo:2018}. Our goal here is less to provide a new calibration, rather then test the validity of current calibrations for our sample. 
Specifically, the lines we compare with each other are H$\alpha$, H$\beta$ and \ion{Mg}{2}. We will discuss correlations with the broad \ion{C}{4} line separately in section~\ref{sec:civ}, since the use of this line as a virial black hole mass estimator is still under debate.

In Figure~\ref{fig:compBalmer}, we show the relation between H$\alpha$ and H$\beta$ FWHM, luminosity and \mbh and the  $L_{\rm{H}\alpha}-L_{5100}$ relation. In addition to our FMOS sample, we show the sample of luminous QSOs from \citet{Shen:2012b}, to cover a wide luminosity range. We compare these to established relations from the literature. For FWHM and  $L_{\rm{H}\alpha}$ vs. $L_{5100}$, we use the empirical relations from \citet{Jun:2015}. These are based on a large compilation of measurements from the literature, augmented by their own measurements at the highest luminosities. They establish these relationships over a wide range in luminosity and are consistent with previous studies over their common luminosity range \citep{Greene:2005,Shen:2012b}. Specifically, these are\footnote{We here denote the units FWHM$_{\rm{H}\beta,3} =$ FWHM$_{\rm{H}\beta}/10^3$~km~s$^{-1}$,  $L_{\rm{H}\alpha,42}=L_{\rm{H}\alpha}/10^{42}$~erg s$^{-1}$ and so forth.}
\begin{equation}
\log \rm{FWHM}_{\rm{H}\beta,3} = 1.061 \log  \rm{FWHM}_{\rm{H}\alpha,3} + 0.055.  \label{eq:w_ba}
\end{equation}
\begin{equation}
\log L_{\rm{H}\alpha,42} = 1.044 \log L_{5100,44} + 0.646. \label{eq:l_ab}
\end{equation}
For the relationship between $L_{\rm{H}\alpha}$ and $L_{\rm{H}\beta}$ we assume as default relation a Balmer decrement of 3.1, as expected for Case B recombination. For the relation $\mbh(\rm{H}\alpha - \rm{H}\beta)$ we assume a one-to-one relation. These default relations are shown as black dashed line in Figure~\ref{fig:compBalmer}. We provide the mean and the standard deviation of our data around these relations in Table~\ref{tab:linecor}. Our data is fully consistent with the reference relationships, with a mean offset of less than 0.04~dex. 
We find for the FMOS sample that H$\beta$ is on average broader than H$\alpha$ by a factor 1.34. This is consistent with previous studies, which focused on low-$z$ AGN \citep{Osterbrock:1982,Greene:2005,Shen:2008,Schulze:2010}. \rev{This indicates a slightly larger virial velocity for H$\beta$, being emitted closer to the black hole. Since H$\beta$ is emitted preferentially in regions of higher density and/or higher ionization parameter \citep{Osterbrock:2006}, this is expected for an increasing density or ionization parameter in the BLR with decreasing radius.}

\rev{In addition, we perform a linear regression on the line correlations. For consistency with the adopted reference relations by \citet{Jun:2015}, we here use the BCES method \citep[][see also \citealt{Nemmen:2012}]{Akritas:1996}, which accounts for measurement errors in both variables and for intrinsic scatter. We also tested the {\tt linmix\_err} method \citep{Kelly:2007}, finding consistent results. We use orthogonal regression, i.e. we do not specify a unique response variable but treat both symetrically, and derive errors via bootstrapping.}
We consider two cases (1) only including our FMOS sample, and (2) adding the luminous QSO sample from \citet{Shen:2012b}. The results are listed in Table~\ref{tab:linecor} and shown in Figure~\ref{fig:compBalmer} as red dashed line and blue solid line respectively. As previously mentioned, we find generally very good agreement between relations based on our data and those reported in the literature over the parameter range covered by our sample. For the $L_{\rm{H}\alpha}-L_{5100}$ correlation based only on the FMOS sample, we find a deviation at the luminous end where our best-fit is extrapolated beyond the luminosity regime of our sample. Including the luminous SDSS QSO sample into the regression, brings our best-fit in good agreement with the reference relation over the full luminosity range ($10^{43}<L_{\rm{H}\alpha}<10^{46}$ erg s$^{-1}$).
We conclude that in their H$\alpha$ and H$\beta$ properties our AGN are fully consistent with the general broad-line AGN population.

\begin{deluxetable*}{ccccccccccc}
\tabletypesize{\scriptsize}
\tablecaption{Line correlations for FWHM, luminosity and \mbh}
\tablewidth{18cm}
\tablehead{
\colhead{Lines} & \colhead{Properties} & \colhead{Sample} & \colhead{$N$} & \colhead{$a_{\rm{ref}}$} & \colhead{$b_{\rm{ref}}$}  & \colhead{$a$} & \colhead{$b$}  & \colhead{$\sigma$} & \colhead{$\Delta X_{\rm{ref}}$} & \colhead{$\sigma_{\rm{ref}}$}\\
}
\startdata
H$\alpha$$-$H$\beta$ & FWHM & FMOS & 35  & $ 1.061$ &  $ 0.055$ & $  1.07\pm  0.23$ &    0.09$\pm  0.12$ & $  0.16$ & $  0.04$ & $  0.16$\\ 
 &  & all & 95  &  &   & $  0.94\pm  0.14$ &    0.12$\pm  0.08$ & $  0.12$ & $ -0.01$ & $  0.12$\\ 
 & $L_{\rm{H}\alpha,42}-L_{5100,44}$ & FMOS & 35  & $ 0.958$ &  $-0.619$ & $  1.12\pm  0.12$ &   -0.91$\pm  0.22$ & $  0.18$ & $ -0.00$ & $  0.17$\\ 
 &  & all & 95  &  &   & $  0.93\pm  0.03$ &   -0.60$\pm  0.08$ & $  0.13$ & $ -0.05$ & $  0.13$\\ 
 & $L_{\rm{H}\alpha,44}-L_{\rm{H}\beta,44}$ & FMOS & 35  & $ 1.000$ &  $-0.491$ & $  1.02\pm  0.12$ &   -0.43$\pm  0.05$ & $  0.19$ & $  0.05$ & $  0.18$\\ 
 &  & all & 95  &  &   & $  0.98\pm  0.03$ &   -0.49$\pm  0.03$ & $  0.17$ & $ -0.00$ & $  0.17$\\ 
 & \mbh & FMOS & 33  & $ 1.000$ &  $ 0.000$ & $  1.13\pm  0.12$ &   -0.04$\pm  0.10$ & $  0.28$ & $  0.06$ & $  0.27$\\ 
 &  & all & 93  &  &   & $  0.90\pm  0.05$ &    0.08$\pm  0.07$ & $  0.22$ & $ -0.04$ & $  0.23$\\ 
H$\alpha$$-$\ion{Mg}{2} & FWHM & FMOS & 126  & $ 0.865$ &  $ 0.115$ & $  0.86\pm  0.09$ &    0.13$\pm  0.05$ & $  0.17$ & $  0.01$ & $  0.17$\\ 
 &  & all & 186  &  &   & $  0.85\pm  0.07$ &    0.13$\pm  0.05$ & $  0.14$ & $  0.00$ & $  0.14$\\ 
 & $L_{\rm{H}\alpha,42}-L_{3000,44}$ & FMOS & 126  & $ 0.932$ &  $-0.315$ & $  1.01\pm  0.07$ &   -0.46$\pm  0.11$ & $  0.26$ & $ -0.04$ & $  0.25$\\ 
 &  & all & 186  &  &   & $  0.89\pm  0.02$ &   -0.30$\pm  0.05$ & $  0.22$ & $ -0.07$ & $  0.23$\\ 
 & \mbh & FMOS & 126  & $ 1.000$ &  $ 0.000$ & $  0.79\pm  0.09$ &    0.11$\pm  0.07$ & $  0.37$ & $ -0.02$ & $  0.42$\\ 
 &  & all & 186  &  &   & $  0.89\pm  0.04$ &    0.07$\pm  0.06$ & $  0.34$ & $ -0.03$ & $  0.36$\\ 
H$\beta$$-$\ion{Mg}{2} & FWHM & FMOS & 36  & $ 0.816$ &  $ 0.064$ & $  1.10\pm  0.85$ &   -0.14$\pm  0.59$ & $  0.22$ & $ -0.00$ & $  0.18$\\ 
 &  & all & 115  &  &   & $  1.10\pm  0.22$ &   -0.14$\pm  0.15$ & $  0.16$ & $ -0.01$ & $  0.13$\\ 
 & $L_{5100,44}-L_{3000,44}$ & FMOS & 36  & $ 0.973$ &  $ 0.287$ & $  1.08\pm  0.14$ &    0.19$\pm  0.14$ & $  0.21$ & $  0.02$ & $  0.21$\\ 
 &  & all & 115  &  &   & $  0.94\pm  0.02$ &    0.31$\pm  0.05$ & $  0.13$ & $ -0.04$ & $  0.14$\\ 
 & \mbh & FMOS & 36  & $ 1.000$ &  $ 0.000$ & $  1.16\pm  0.36$ &   -0.15$\pm  0.33$ & $  0.51$ & $ -0.01$ & $  0.47$\\ 
 &  & all & 115  &  &   & $  1.11\pm  0.10$ &   -0.11$\pm  0.15$ & $  0.35$ & $  0.04$ & $  0.33$\\ 
H$\alpha$$-$\ion{C}{4} & FWHM & FMOS & 28  & $ 1.007$ &  $ 0.031$ & $ -1.08\pm  1.60$ &    1.14$\pm  0.84$ & $  0.20$ & $  0.04$ & $  0.28$\\ 
 &  & all & 88  &  &   & $  1.05\pm  0.63$ &    0.05$\pm  0.18$ & $  0.21$ & $  0.05$ & $  0.21$\\ 
 & $L_{\rm{H}\alpha,42}-L_{1350,44}$ & FMOS & 28  & $ 0.933$ &  $-0.213$ & $  0.58\pm  0.52$ &    0.53$\pm  0.89$ & $  0.34$ & $  0.15$ & $  0.30$\\ 
 &  & all & 88  &  &   & $  0.77\pm  0.12$ &    0.26$\pm  0.31$ & $  0.26$ & $  0.06$ & $  0.27$\\ 
 & \mbh & FMOS & 28  & $ 1.000$ &  $ 0.000$ & $  0.19\pm0.94$ &   0.52$\pm0.57$ & $  0.43$ & $  0.04$ & $  0.56$\\ 
 &  & all & 88  &  &   & $  0.92\pm  0.15$ &    0.10$\pm  0.20$ & $  0.41$ & $ -0.00$ & $  0.43$\\ 
H$\beta$$-$\ion{C}{4} & FWHM & FMOS & 30  & $ 0.949$ &  $-0.023$ & $ -0.61\pm 0.88$ &    0.99$\pm 0.59$ & $  0.24$ & $ -0.02$ & $  0.30$\\ 
 &  & all & 113  &  &   & $  0.34\pm 0.75$ &    0.45$\pm 0.52$ & $  0.17$ & $  0.05$ & $  0.22$\\ 
 & $L_{5100,44}-L_{1350,44}$ & FMOS & 30  & $ 0.974$ &  $ 0.391$ & $  0.59\pm  0.51$ &    0.91$\pm  0.69$ & $  0.29$ & $  0.03$ & $  0.31$\\ 
 &  & all & 113  &  &   & $  0.90\pm  0.12$ &    0.60$\pm  0.25$ & $  0.22$ & $  0.05$ & $  0.23$\\ 
 & \mbh & FMOS & 30  & $ 1.000$ &  $ 0.000$ & $  0.47\pm  2.50$ &    0.29$\pm  2.04$ & $  0.43$ & $ -0.17$ & $  0.58$\\ 
 &  & all & 113  &  &   & $  1.18\pm  0.22$ &   -0.24$\pm  0.30$ & $  0.50$ & $  0.00$ & $  0.46$\\ 
\enddata
\tablecomments{BCES othogonal regression result to the relation $\log Y = a \log X  + b$ to the line and continuum measurements of several broad-lines $X$ and $Y$, namely H$\alpha$, H$\beta$ and \ion{Mg}{2}. We also list the scatter around the best-fit relation as $\sigma$ and the parameters for a reference relation as discussed in the text and provide the mean and the standard deviation of our sample around this reference relation as $\Delta X_{\rm{ref}}$ and $\sigma_{\rm{ref}}$ respectively.}
\label{tab:linecor}
\end{deluxetable*}

In Figure~\ref{fig:compMgII}, we show the correlations of FWHM, luminosity and black hole mass for H$\alpha$ and H$\beta$ with \ion{Mg}{2} for those that have \ion{Mg}{2} measurements from optical spectroscopy. We base our reference relation for FWHM and luminosity again on the study by \citet{Jun:2015}, which reports
\begin{equation}
\log \rm{FWHM}_{\rm{H}\beta,3} = 1.226 \log  \rm{FWHM}_{\rm{MgII},3} + 0.078
\end{equation}
\begin{equation}
\log L_{3000,44} = 0.973 \log L_{5100,44} + 0.287
\end{equation}
The reference relation for H$\alpha$ against \ion{Mg}{2} is obtained by combining these with Equations~\ref{eq:w_ba} and \ref{eq:l_ab}. For \mbh, we again assume a one-to-one relation. Our FMOS sample is fully consistent with those reference relationships. We give their mean and dispersion around the reference relations in Table~\ref{tab:linecor}. Additionally, we provide the best-fit linear regression result based on a fit to the FMOS sample only as well as to the combination of the FMOS sample with the high luminosity QSO samples from \citet{Shen:2012b} and \citet{Zuo:2015}\footnote{The sample by \citet{Zuo:2015} is only used for  H$\beta$ vs. \ion{Mg}{2} as they do not observe H$\alpha$}. We find an excellent agreement with the reference relations. 

Our sample of moderate-luminosity AGN at high redshift is fully consistent with previous work, which mainly combined luminous high-$z$ QSOs with moderate-luminosity low-$z$ AGN. We conclude that the relationship of \ion{Mg}{2} with the Balmer lines is fully consistent with the broad-line AGN population, typically observed at lower redshift.

\begin{figure*}
\centering
\includegraphics[width=18cm,clip]{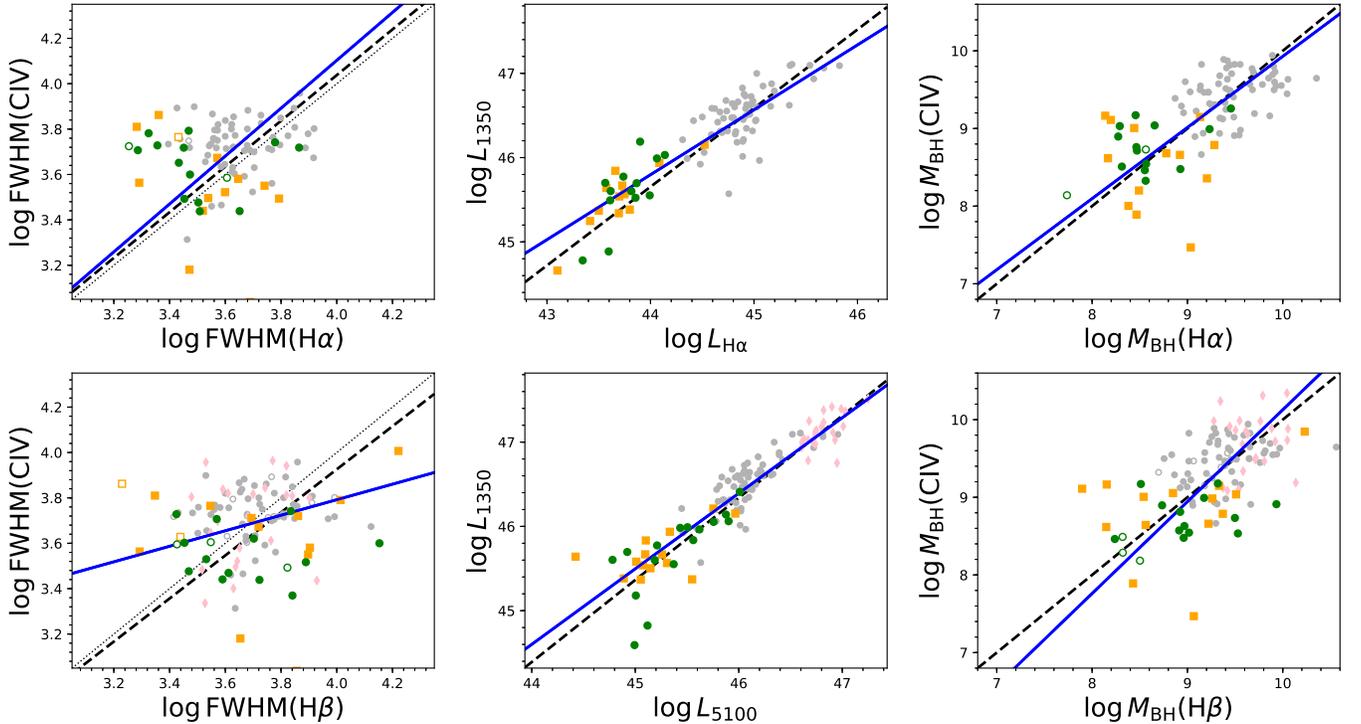} 
\caption{Comparison of  spectral measurements and derived black hole masses between H$\alpha$/H$\beta$  and \ion{C}{4} for those AGN with \ion{C}{4} measurements from optical spectra in COSMOS (green squares) and SXDS (orange circles), augmented by the luminous quasar samples by \citet[][gray circles]{Shen:2012b} and \citet[][pink diamonds; \ion{C}{4} measurements taken from \citet{Shen:2011}]{Zuo:2015}. Lines are as in Figure~\ref{fig:compMgII}.
}
\label{fig:compCiv} 
\end{figure*}

\subsection{\ion{C}{4} line correlations} \label{sec:civ}
The broad \ion{C}{4} line is sometimes used as a virial black hole mass estimator, enabling black hole mass estimates at $2<z<5$ from optical spectroscopy \citep[e.g.][]{Vestergaard:2004,Vestergaard:2009,Shen:2011,Kelly:2013}. However, its reliability is questionable, since it is often severely affected by a non-virial component of the BLR gas \citep[][and references therein]{Baskin:2005, Trakhtenbrot:2012,Denney:2012,Coatman:2017}. This potentially outflowing component is especially significant for the most luminous quasars observed at high redshift. 
However, the \ion{C}{4} line appears to be a more robust black hole mass estimator for local low-luminosity AGNs \citep{Vestergaard:2006,Tilton:2013}. Therefore, it is worth reexamining the reliability of \ion{C}{4} as virial black hole mass estimator especially for moderate-luminosity AGN at high-$z$. 

For our FMOS sample in the COSMOS and SXDS fields, we obtain optical spectra from SDSS \citep{Abazajian:2009}, BOSS \citep{Alam:2015} and zCOSMOS-Deep \citep{Lilly:2007}. We measure the \ion{C}{4} line for 43 AGN (25 in COSMOS and 18 in SXDS). We have excluded CID~346 here due to the presence of a strong \ion{C}{4} BAL, preventing a robust measurement of the intrinsic line profile. We tied the absolute flux calibration to the optical photometry from Subaru \citep{Laigle:2016,Mehta:2018}.

We fit the \ion{C}{4} spectral region following previous studies \citep[e.g.][]{Shen:2011}. For each spectrum, we first fit the local continuum by a power-law. The broad \ion{C}{4} line is fit using up to three Gaussian components over the interval $1450-1700$\AA{}.  In addition, we allow for the inclusion of the \ion{He}{2}$\lambda 1640$, \ion{O}{3}]$\lambda1663$ and \ion{N}{4}]$\lambda1486$ lines, each modeled by a single broad Gaussian component, with a common line width and velocity shift \citep{Fine:2010,Trakhtenbrot:2012}. We manually mask out spectral regions affected by narrow absorption features which can affect the line fit.
We obtained the \ion{C}{4} FWHM and the continuum luminosity at 1350\AA{}, $L_{1350}$, from the best-fit model and their uncertainties from Monte Carlo simulations, as we did for the other emission line measurements (see section~\ref{sec:line}). To estimate \mbh, we use the virial relation for \ion{C}{4} by \citet{Vestergaard:2006}.

In Figure~\ref{fig:compCiv}, we compare the FWHM and luminosity measurements and the \mbh estimates from \ion{C}{4} to those from the Balmer lines measured from the FMOS spectra. We again compare these with the relations given by \citet{Jun:2015}
\begin{equation}
\log \rm{FWHM}_{\rm{H}\beta,3} = 1.054 \log  \rm{FWHM}_{\rm{CIV},3} + 0.024. \label{eq:fwhm_civ}
\end{equation}
\begin{equation}
\log L_{1350,44} = 0.974 \log L_{5100,44} + 0.391 \ ,  \label{eq:l1350}
\end{equation}
shown as black dashed line. Furthermore, we show the best-fit relation to the FMOS sample, combined with the luminous QSOs from \citet{Shen:2012b} and \citet{Zuo:2015} by the blue solid line. Their best-fit values are given in Table~\ref{tab:linecor}. 
The luminosity $L_{1350}$ shows a good correlation with both $L_{5100}$ and $L_{\rm{H}\alpha}$, consistent with the reference relation given in Equation~\ref{eq:l1350}. 
The FWHM measurements show a large scatter between \ion{C}{4} and the Balmer lines, with $\sigma\sim0.3$~dex. A Spearman rank-order test does not find a statistically significant correlation between the FWHM of \ion{C}{4} and the Balmer lines for our sample. While measurement uncertainties due to the data quality of both the optical and near-IR spectra will have a significant contribution to this scatter, it is however clear that the \ion{C}{4} FWHM has a significantly weaker correlation with the FWHM of the Balmer lines than \ion{Mg}{2}. 

Under the assumption of virialized motion we would expect FWHM(\ion{C}{4}) $ > $ FWHM(H$\alpha$ or H$\beta$). However, this is not seen in our sample, consistent with several previous studies \citep[][and references therein]{Trakhtenbrot:2012} and with the reference relation given by Equation~\ref{eq:fwhm_civ}. We find FWHM(\ion{C}{4}) $<$ FWHM(H$\alpha$) for 50\% (14/28) and FWHM(\ion{C}{4}) $<$ FWHM(H$\beta$) for 60\% (18/30) of our sources. As discussed in \citet{Trakhtenbrot:2012}, this suggest that \ion{C}{4} is not virialized, as is required for the application of the virial method.
Thus, our results support the notion of \ion{C}{4} as a less reliable black hole mass estimator, suffering from large uncertainties.
This is confirmed by the direct comparison of the \mbh estimates in the right panels of Figure~\ref{fig:compCiv}. However, while there is a significant scatter of $0.43-0.48$~dex in the \mbh estimates, the combined sample from FMOS and the luminous QSOs on average is fully consistent with the one-to-one relation between \mbh(\ion{C}{4}) and the \mbh estimate from the Balmer lines.

\begin{figure*}
\centering
\includegraphics[width=18cm,clip]{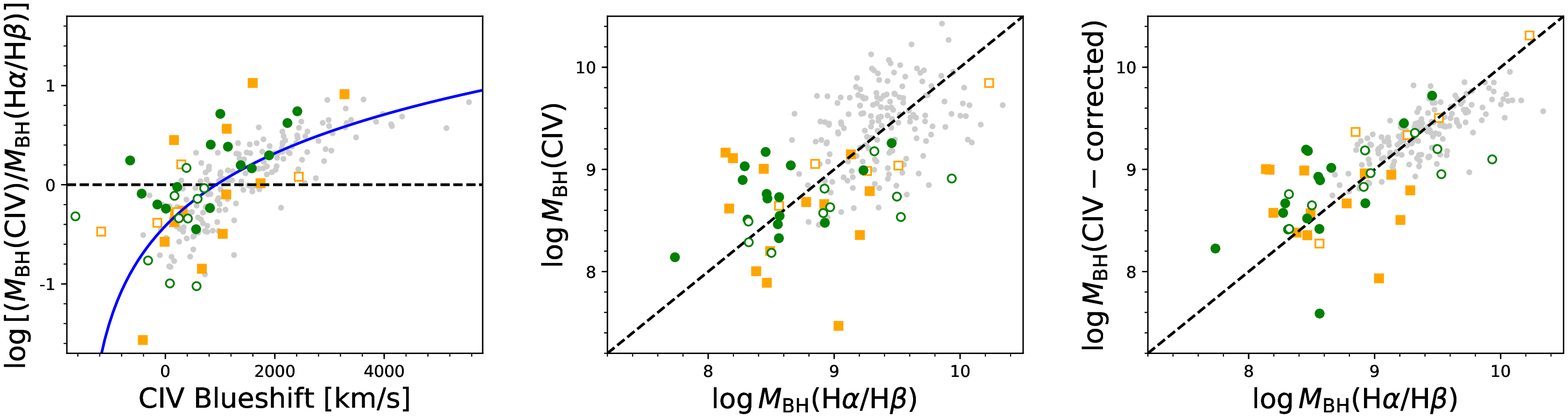} 
\caption{Left panel: Ratio of SMBH mass estimates from \ion{C}{4} and the Balmer lines against the \ion{C}{4} blueshift, measured in respect to either the peak of the Balmer lines or of [\ion{O}{3}] in the FMOS spectra. Filled green circles (COSMOS) and orange squares (SXDS) are for H$\alpha$ based \mbh, while the open symbols are for H$\beta$ based \mbh. The light gray circles show the sample by \citet{Coatman:2017} and the blue solid line is their best-fit relation for their sample. Central panel: Comparison of \ion{C}{4} and H$\alpha$/H$\beta$ based \mbh using the virial formula by \citet{Vestergaard:2006}. The black dashed line shows the one-to-one relation. Right panel: Same for \ion{C}{4}  \mbh estimates using the blueshift based correction from \citet{Coatman:2017}.
}
\label{fig:compCivMBH} 
\end{figure*}

There have been several attempts to improve the reliability of \ion{C}{4} based SMBH masses \citep{Denney:2012,Runnoe:2013,Park:2013,Coatman:2017}. The study by \citet{Coatman:2017} demonstrated a correlation between the FWHM ratio of \ion{C}{4} to the Balmer lines and the \ion{C}{4} blueshift for very luminous QSOs. They propose an improved \ion{C}{4} virial \mbh estimator, including knowledge of the  \ion{C}{4} blueshift (but see \citealt{Mejia:2018b} for a contradictory argument). We here test if (1) our moderate-luminosity AGN sample follows the same correlation, and (2) if their improved \mbh estimator also provides a significant improvement for these moderate-luminosity AGN. The \ion{C}{4} blueshift for our FMOS sample is defined as the velocity offset of the peak of the \ion{C}{4} line, measured from the best-fit, in respect to the near-IR redshift, as discussed in section~\ref{sec:line}.

In the left panel of Figure~\ref{fig:compCivMBH}, we plot the ratio \mbh(\ion{C}{4})/\mbh(H$\alpha$ or H$\beta$) against the \ion{C}{4} blueshift.
For comparison we also show the luminous QSO sample from \citet{Coatman:2017}  as gray circles and their best-fit relation as solid blue line. While our sample shows a larger scatter, it is consistent with the relation found by \citet{Coatman:2017}. However, we note that our moderate-luminosity AGN sample shows a narrower distribution of \ion{C}{4} blueshifts, largely lacking very strong blueshifts, with a median velocity shift of 820 km/s compared to 1290 km/s. This can be understood as a consequence of the lower luminosities of the FMOS sample. Less luminous AGN have on average higher \ion{C}{4} equivalent width \citep[the well known Baldwin effect,][]{Baldwin:1977} and high \ion{C}{4} EW AGN tend to show a lack of large blueshifts \citep{Richards:2011}.

In the central and right panel of Figure~\ref{fig:compCivMBH}, we show the comparison of the Balmer line \mbh estimates to the \ion{C}{4} line \mbh estimates based on \citet{Vestergaard:2006} and based on the blueshift-based correction prescription from \citet{Coatman:2017} respectively.  We find that the \citet{Coatman:2017} prescription provides an improvement on the \ion{C}{4} masses, with the dispersion around the one-to-one relation decreasing from 0.53 to 0.43~dex for the FMOS sample (while the mean offset changes from $-0.09$ to $0.06$). However, this is less than the improvement from 0.4 to 0.2 dex found for the luminous QSOs in \citet{Coatman:2017}. We attribute at least part of this reduced effectiveness to the lack of large blueshift AGN ($>3000$~km s$^{-1}$) in our moderate-luminosity AGN sample. For these objects, the improvement is most significant as illustrated in the left panel of Figure~\ref{fig:compCivMBH}. However, we caution that the spectra for our sample are typically of lower quality in both the optical and NIR than those used in \citet{Coatman:2017}.

We conclude that moderate-luminosity AGN follow a similar relation between their \ion{C}{4} and Balmer FWHM ratio and the \ion{C}{4} blueshift as luminous AGN, but are lacking the highest blueshift objects as a consequence of  the Baldwin effect. When we apply a correction to the virial formula using the \ion{C}{4} blueshift, this leads to an improvement of the \mbh estimates. However, the deficit of high blueshift sources somewhat reduces the overall importance and effectiveness of such a correction for moderate-luminosity AGN like those studied here, compared to the most luminous QSOs.

\section{Discussion} \label{sec:discuss}
\subsection{The virial black hole mass estimators}
In the previous sections, we presented a comparison of broad-line and continuum measurements as well as the resulting black hole mass estimates for different AGN broad emission lines, commonly used for the virial method. Our FMOS study is unique in that it robustly tests these relationships and the reliability of current calibrations of the virial method for an unprecedentedly large sample of moderate-luminosity AGN at high redshift, that is probing the bulk of the (unobscured) AGN population at the epoch of peak SMBH mass assembly. As shown in section~\ref{sec:linecomp}, the empirical relationships in  FWHM and luminosity between H$\alpha$, H$\beta$ and \ion{Mg}{2} hold for moderate-luminosity AGN at $z\gtrsim1$, consistent with previous work based on smaller samples \citep[M13][]{Karouzos:2015,Suh:2015}. Furthermore, the virial \mbh estimators presented in section~\ref{sec:mbh} provide consistent results.

For the correlation between H$\alpha$ and H$\beta$ \mbh, we find a scatter of $\sim0.3$~dex, which is similar to our estimate of the measurement uncertainty in section~\ref{sec:LRHR} due to data quality. This confirms the good agreement between these two black hole mass estimators.  Both lines provide statistically equivalent black hole mass estimates.

The \mbh estimates  based on \ion{Mg}{2} show a larger scatter to the Balmer lines based \mbh of $\sim0.4-0.5$~dex, but no statistically significant offset when consistently calibrated virial relationships are used. Part of the increased scatter is caused by the non-simultaneous nature of the optical and near-IR spectra, i.e. the effect of AGN variability, and the typical rather low S/N in both of them. However, it has been shown in several studies that \ion{Mg}{2} based \mbh show a larger scatter to Balmer line based \mbh than those among the Balmer line calibrations itself \citep[e.g.][]{Shen:2012b,Mejia:2016}, qualitatively consistent with our results.

\begin{figure*}
\centering
\includegraphics[width=8.8cm,clip]{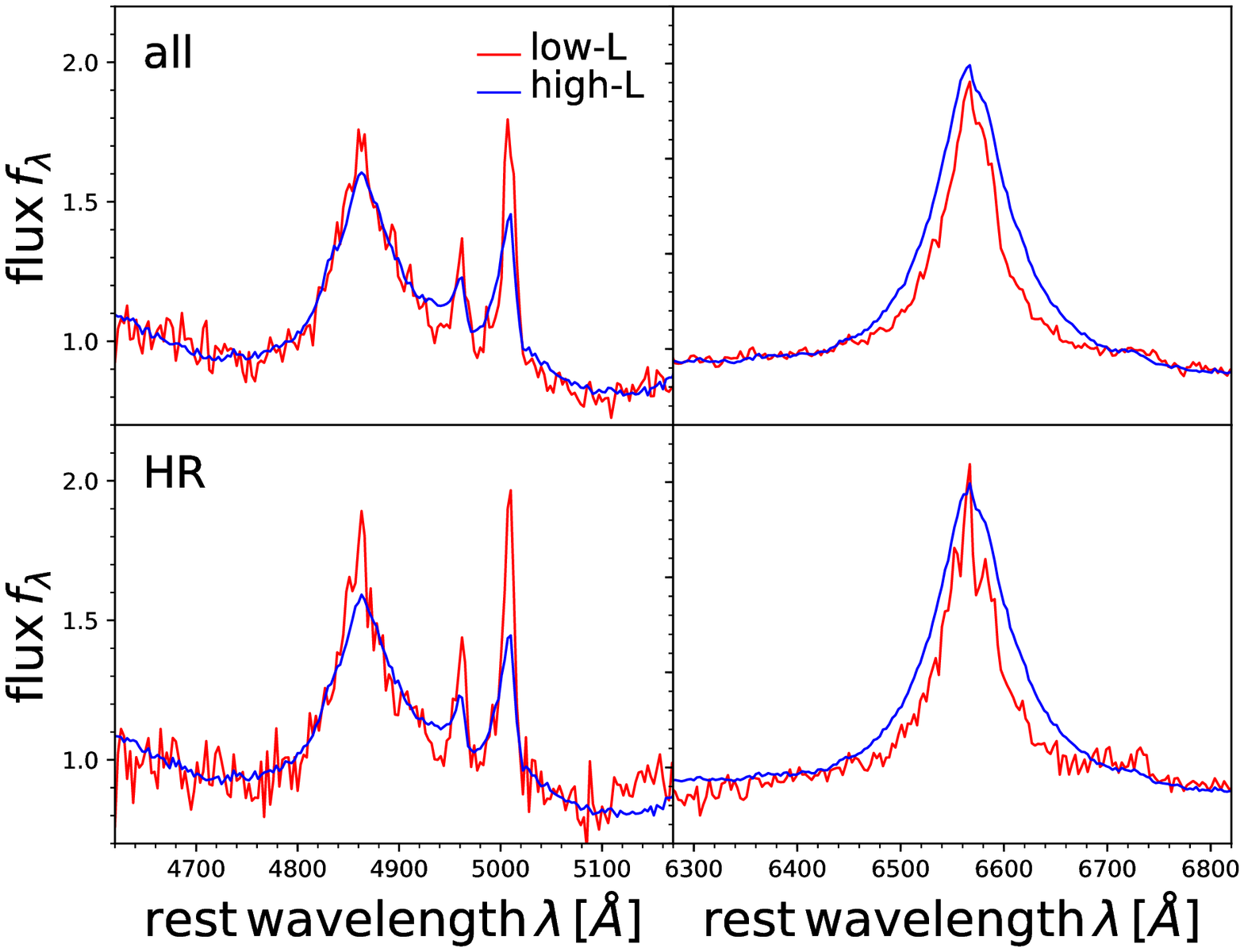} 
\includegraphics[width=8.8cm,clip]{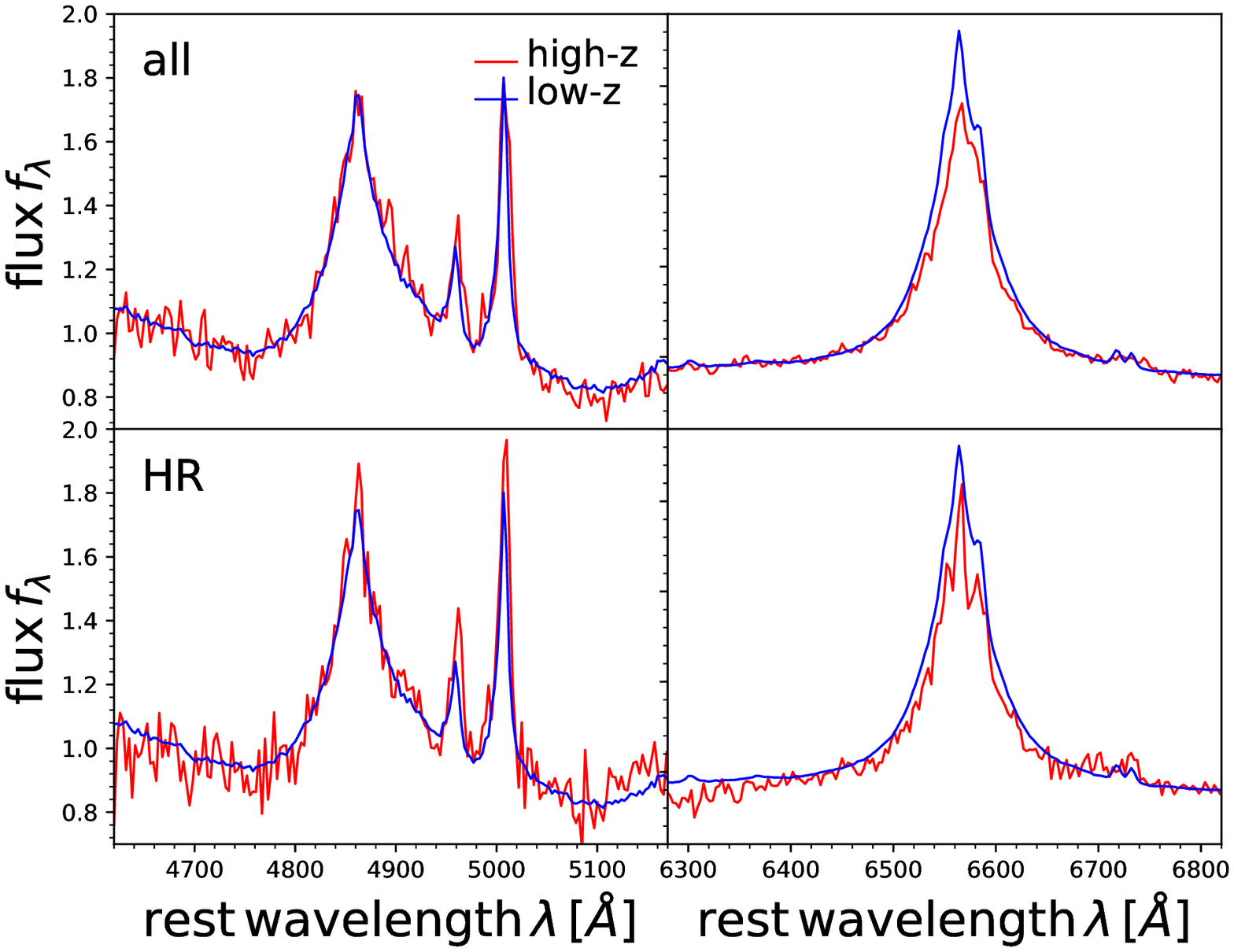}
\caption{Comparison of the stacked spectra for the FMOS sample with broad H$\alpha$ line detection and broad H$\beta$ line detection (red line) with two comparison samples (blue line). The left panel shows the comparison with 
with the stacked spectra of a sample of luminous quasars at similar redshift. The right panel shows the comparison with a low redshift quasar sample matched in luminosity. In the upper panels we have used all FMOS spectra (LR and HR mode), while in the lower panels only the COSMOS spectra in HR mode are used to produce the composite.}
\label{fig:stack} 
\end{figure*}

The \mbh estimates  based on \ion{C}{4} show a larger scatter than those based on \ion{Mg}{2}, but also no statistically significant offset when using consistent local virial calibrations.
This is consistent with the common understanding that \ion{C}{4} based \mbh estimates bear large uncertainties. Correcting these \mbh estimates by using \ion{C}{4} blueshifts, as suggested by \citet{Coatman:2017}, improves the agreement with the Balmer line based \mbh, leading to a comparable scatter as what we find for the \ion{Mg}{2} based \mbh. While higher S/N data would be needed to draw more robust conclusions this seems to support the notion that it is possible to rehabilitate the use of \ion{C}{4} as a virial SMBH mass estimator \citep{Runnoe:2013,Mejia:2016,Coatman:2017}.

Our results provide validation for the use of H$\alpha$, H$\beta$ and  \ion{Mg}{2} for moderate-luminosity AGN out to high-$z$. However, we caution that the relative precision of these three black hole mass estimators does not inform us about their overall accuracy. The single-epoch method to estimate black hole masses is most likely prone to systematics which are unaccounted for and are often hard to quantify \citep[e.g.][]{Shen:2013}, including the inclination of the BLR towards our line of sight \citep{Collin:2006,Decarli:2008,Mejia:2018}, radiation pressure effects \citep{Marconi:2008,Netzer:2010}, etc. Larger samples of AGN with \mbh measurements complimentary to virial \mbh estimates are crucial to improve on the accuracy of the virial method as an effective mean to infer the mass of a SMBH. This can include direct dynamical measurements \citep{Onken:2007}, large scale reverberation mapping campaigns \citep{Peterson:2004,Shen:2015,Kollmeier:2017} or accretion disc modeling  \citep{Capellupo:2016}.

\subsection{Redshift and luminosity dependence of rest-frame optical spectral properties}
Our sample enables us to disentangle redshift evolution and luminosity effects in the rest-frame optical spectra of broad-line AGN. While moderate-luminosity AGN are easily probed at low $z$, the space density of luminous quasars is too low  to enclose a sizable sample of luminous quasars in the local volume. On the other hand, at $z>1$ luminous quasars are much more common and can easily be detected by large area surveys \citep{Richards:2006b}, and their near-IR follow up can be carried out at moderate size telescopes. Observations of the rest-frame optical spectra of moderate-luminosity AGN are more demanding and require 8m-class telescopes. This makes a one-to-one comparison of luminosity matched samples at different redshifts often challenging.

Here, we compare the composite spectra for our moderate-luminosity AGN sample with a luminous AGN sample at comparable redshift and with a low-$z$ AGN sample at matched luminosity. We stack the spectra of our FMOS sample with H$\alpha$ (211 AGN) and H$\beta$ detection (63 AGN) separately.
For the luminous AGN comparison sample, we use the study by \citet{Shen:2012b}\footnote{The reduced spectra have been kindly made publicly available in \citet{Shen:2016}.}, consisting of 60 quasars with  H$\alpha$ and H$\beta$ coverage. For the low-$z$ sample, we construct a luminosity-matched sample at $z<0.84$ (for H$\beta$) or $z<0.35$ (for H$\alpha$) from the SDSS DR7 quasar catalog \citep{Schneider:2010,Shen:2011}. For every FMOS AGN, we find the three closest luminosity matches in $L_{5100}$ (for H$\beta$) or in $L_{\rm{H}\alpha}$ (for H$\alpha$).

Each spectrum is shifted into rest-frame (based on the near-IR redshift), re-binned to a common wavelength scale and normalized at 5100\AA{} for the H$\beta$ stack and at 6400\AA{} for the H$\alpha$ stack. 
Stacked spectra are then generated using the median. Uncertainties are derived from bootstrapping the sample where we applied the Monte-Carlo approach discussed in Section~\ref{sec:line} to every bootstrapped object. In addition to generating a composite spectrum for the full FMOS sample, we also only use the HR spectra, in which case we maintain the higher resolution during the stacking process.

In the left panel of Figure~\ref{fig:stack}, we show the composite spectra for the FMOS sample, compared to the luminous quasar sample from \citet{Shen:2012b}.  For the H$\alpha$ line, the most prominent difference is the narrower width of the broad-line. This is consistent with the FWHM$_{\rm{H}\alpha}$ distribution of the two samples, with a median FWHM$_{\rm{H}\alpha}$ of 3550 km~s$^{-1}$ for the FMOS sample and 4233 km~s$^{-1}$ for the luminous quasar sample. The latter value is mainly driven by the lack of FWHM$_{\rm{H}\alpha}<2500$ km~s$^{-1}$ in the luminous quasar sample. This is a physical consequence of the Eddington limit which is largely obeyed in both samples. A maximum \er\ set by the Eddington limit translates into a minimum FWHM at a given luminosity, which particularly restricts the possible range in FWHM for the most luminous quasars. The broad H$\beta$ line does not show as pronounced a difference with median FWHM$_{\rm{H}\beta}\sim5000$ km~s$^{-1}$ in both samples. This can be understood since the H$\beta$ FMOS sample has a higher average luminosity than the H$\alpha$ sample and thus the luminosity difference to the high-$L$ comparison sample is smaller, as for example shown in Figure~\ref{fig:me}. We do see a stronger prominence of the narrow lines in the FMOS sample in both H$\alpha$ and H$\beta$. 
The most prominent difference in the H$\beta$ region is the strength and profile of  [\ion{O}{3}]. The moderate-luminosity sample from FMOS has a significantly larger [\ion{O}{3}] equivalent width (EW) than the luminous quasar sample. This Baldwin effect type behavior of the [\ion{O}{3}] EW is well known, especially at $z<1$ \citep{Stern:2013,Zhang:2013,Shen:2014} and for luminous quasars ar $z>2$ \citep{Netzer:2004}. Here, we extend the trend to significantly lower luminosities at $z>1$. We also find evidence for a more prominent blue wing component of the  [\ion{O}{3}] line in the high-$L$ composite, indicative of a higher strength and/or ubiquity of ionized outflows in luminous quasars, consistent with previous work  \citep{Shen:2014,Bischetti:2017}.

In the right panel of Figure~\ref{fig:stack}, we show the comparison of our high-$z$ sample with the lower-$z$ match from the SDSS quasar catalog. For the H$\beta$ region, we find the higher-$z$ sample to be fully consistent with the lower-$z$ matched sample. This indicates no significant redshift evolution in the average rest-frame optical properties of luminosity-matched AGN, including broad  H$\beta$, [\ion{O}{3}]  and \ion{Fe}{2}. In the H$\alpha$ region, we find a generally good agreement in the broad H$\alpha$ shape, but when normalized in continuum luminosity the FMOS sample shows weaker broad H$\alpha$. This corresponds to a lower  broad H$\alpha$ EW, consistent with the measurements from the individual fits, where we find median values of EW$_{\rm{H}\alpha}=257$ and 374~km s$^{-1}$ for the high-$z$ and low-$z$ sample respectively.  Potential reasons for this deviation are differences in the host galaxy contribution, sample selection effects (e.g. optical vs. X-ray selection) or intrinsic redshift evolution effects. 


\begin{figure*}
\centering
\includegraphics[width=18cm,clip]{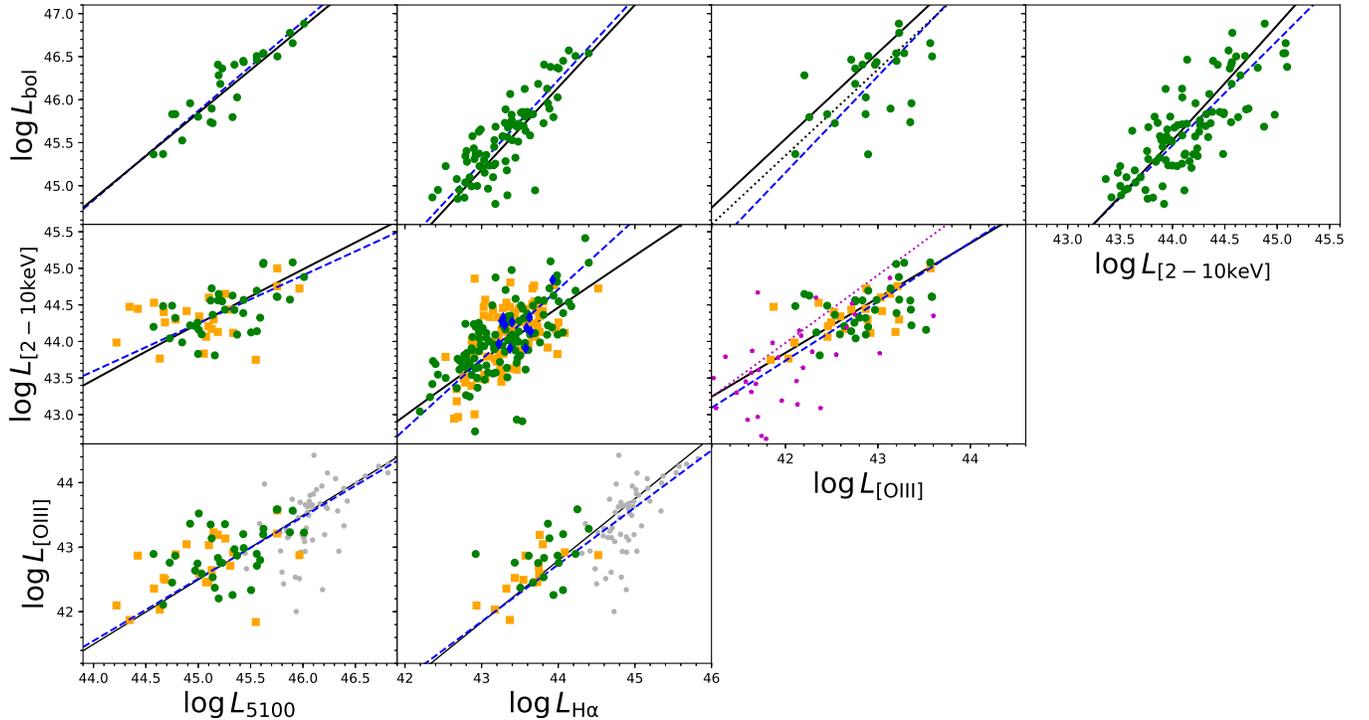} 
\caption{Correlation of various AGN luminosities against each other, namely $L_{\rm{bol}}$, \Lx, $L_{\rm{[OIII]}}$, $L_{5100}$ and $L_{\rm{H}\alpha}$. The color coding for the FMOS sub-samples is as in the previous figures. The best-fit relation to our data is shown as blue dashed line, while the black solid line shows a reference relation, as discussed in the text. In the bottom row, we show the luminous quasar sample by \citet{Shen:2012b} as gray circles and the local sample of SDSS DR7 quasars \citep{Shen:2011} by the gray contours. In the $L_{\rm{[OIII]}}-L_{\rm{bol}}$ panel, the black dotted line shows the expected relation assuming $L_{\mathrm{bol}}/L_{5100}=7$ and $L_{5100}/L_{\rm{[OIII]}}=320$. In the $L_{\rm{[OIII]}}-\Lx$ panel the doted magenta line indicates the relation from \citet{Ueda:2015} and the magenta stars show the type-1 AGN in their sample. }
\label{fig:Lrelation} 
\end{figure*}

\subsection{Correlation of bolometric and X-ray luminosity with optical luminosity indicators} \label{sec:lumcor}
Knowledge of the bolometric luminosity of an AGN is fundamental to understand its  current accretion and growth phase, energy output and impact on its environment. However, in most cases, observations are only available at particular wavelengths and the full SED is often difficult to assess. Thus typically one has to rely on either optical emission line or continuum luminosities or X-ray luminosities as bolometric luminosity indicators. In principle, the optical continuum luminosity traces the accretion disk emission most directly with scatter mainly caused by the variation of intrinsic SED shapes. However, the optical continuum luminosity can suffer from severe contamination by the host galaxy, extinction or will be completely absorbed in the case of obscured AGN. Alternative common bolometric luminosity indicators (especially for obscured AGN) involve hard X-ray luminosity \citep[e.g.][]{Marconi:2004,Vasudevan:2009,Lusso:2012} or [\ion{O}{3}] luminosity \citep[e.g][]{Heckman:2004,Kauffmann:2009,LaMassa:2009,Pennell:2017}. Here, we compare several bolometric luminosity indicators for our sample. In addition, for part of the COSMOS sample, we incorporate bolometric luminosities  presented in \citet{Lusso:2012}. They derived bolometric luminosities for AGN detected within XMM-COSMOS by integrating the observed SED for type~1 AGN from 1 $\mu m$ to 200~keV. In total, 95/141 FMOS AGN in COSMOS with H$\alpha$ or H$\beta$ detection have an $L_{\rm{bol}}$ measurement from \citet{Lusso:2012}. Our sample provides a unique opportunity to study several different  bolometric luminosity indicators in relation to directly integrated $L_{\rm{bol}}$ measurements for a homogeneous sample of AGN at high-$z$.

In Figure~\ref{fig:Lrelation}, we compare measurements of $L_{\rm{bol}}$, \Lx, $L_{\rm{[OIII]}}$, $L_{5100}$ and $L_{\rm{H}\alpha}$ for our FMOS sample with each other. The latter three luminosities are obtained from the FMOS spectra as discussed above, $L_{\rm{bol}}$ for COSMOS AGN is taken from \citet{Lusso:2012} and \Lx\ has been presented in \citet{Civano:2016}, \citet{Akiyama:2015} and \citet{Xue:2011} for COSMOS, SXDS and E-CDF-S respectively. We perform a linear regression using the BCES method for each luminosity comparison, shown by the blue dashed line and listed in Table~\ref{tab:lumcor}. In addition we show a relation taken from the literature for the specific luminosity comparison as black solid line.
We also inspected the flux$-$flux relation for each of the $L-L$ relations we consider here and performed a BCES regression for these. We confirmed that none of the trends discussed here is seriously affected or driven by the $d_L^2$ term that goes into both axes in a $L-L$ plot.

\begin{deluxetable}{cccccc}
\tabletypesize{\scriptsize}
\tablecaption{Luminosity correlations}
\tablewidth{8cm}
\tablehead{
\colhead{$Y$} & \colhead{$X$} & \colhead{$N$}  & \colhead{$a$} & \colhead{$b$}  & \colhead{$\sigma$} 
}
\startdata
$L_{\rm{bol}}$ & $L_{\rm{X}}$ & 95  & $  1.22\pm  0.10$ &  $  1.47\pm  0.04$ & $  0.34$ \\  \noalign{\smallskip} 
$L_{\rm{bol}}$ & $L_{\rm{[OIII]}}$ & 25  & $  1.14\pm  0.20$ &  $  3.41\pm  0.23$ & $  0.43$ \\  \noalign{\smallskip} 
$L_{\rm{bol}}$ & $L_{5100}$ & 27  & $  1.04\pm  0.08$ &  $  0.83\pm  0.12$ & $  0.18$ \\  \noalign{\smallskip} 
$L_{\rm{bol}}$ & $L_{\rm{H}\alpha}$ & 82  & $  0.96\pm  0.07$ &  $  2.23\pm  0.05$ & $  0.26$ \\  \noalign{\smallskip} 
$L_{\rm{X}}$ & $L_{\rm{[OIII]}}$ & 60  & $  0.81\pm  0.16$ &  $  1.35\pm  0.20$ & $  0.32$ \\  \noalign{\smallskip} 
$L_{\rm{X}}$ & $L_{5100}$ & 63  & $  0.66\pm  0.17$ &  $ -0.41\pm  0.21$ & $  0.32$ \\  \noalign{\smallskip} 
$L_{\rm{X}}$ & $L_{\rm{H}\alpha}$ & 204  & $  0.96\pm  0.08$ &  $  0.71\pm  0.06$ & $  0.35$ \\  \noalign{\smallskip} 
$L_{\rm{[OIII]}}$ & $L_{5100}$ & 119  & $  0.96\pm  0.08$ &  $ -2.45\pm  0.15$ & $  0.44$ \\  \noalign{\smallskip} 
$L_{\rm{[OIII]}}$ & $L_{\rm{H}\alpha}$ & 93  & $  0.88\pm  0.07$ &  $ -1.27\pm  0.06$ & $  0.41$ \\  \noalign{\smallskip} 
$L_{5100}$ & $L_{\rm{H}\alpha}$ & 95  & $  0.93\pm  0.03$ &  $  1.26\pm  0.02$ & $  0.13$ \\  \noalign{\smallskip} 
\enddata
\tablecomments{BCES othogonal best-fit to the relation $\log Y = a (\log X - 44) + b + 44$. The correlations between $L_{\rm{H}\alpha}$, $L_{5100}$ and $L_{\rm{[OIII]}}$ include the sample of luminous quasars by \citet{Shen:2012b}.}
\label{tab:lumcor}
\end{deluxetable}

The first row in Figure~\ref{fig:Lrelation} provides the comparison between $L_{\rm{bol}}$ and four common bolometric luminosity indicators. As expected, the best correlation with $L_{\rm{bol}}$ is $L_{5100}$, with a scatter of 0.19~dex. $L_{\rm{H}\alpha}$ is also a good bolometric luminosity indicator for our sample, with a scatter of 0.27~dex. The black solid line in both panels shows the bolometric correction factors  adopted in section~\ref{sec:mbh}, i.e. based on BC$_{5100}=7$. Our best-fit is in good agreement with this assumption, verifying the average robustness of our derived $L_{\rm{bol}}$ and \er\ for the remainder of the sample without direct $L_{\rm{bol}}$ measurement.

Both $L_{\rm{[OIII]}}$ and \Lx\ show a clear correlation with $L_{\rm{bol}}$, but with considerably larger scatter (0.43 and 0.34~dex respectively). However, we note that for the $L_{\rm{[OIII]}}-L_{\rm{bol}}$ correlation the sample size is small.
The [\ion{O}{3}]$\lambda5007$ line is ionized by the extreme-UV part of the accretion disk emission and therefore can serve as an average bolometric luminosity indicator. However, the line strength is also strongly affected by the size and physical conditions in the Narrow Line Region (NLR) where it is emitted from, including its density, clumpiness and the NLR covering factor \citep{Baskin:2005b,Stern:2012}. Furthermore, the emission line is sensitive to extinction by dust along the line of sight. A further source of scatter is the extended spatial scales of the NLR on which [\ion{O}{3}] is emitted, compared to the central engine. Given the time variability of AGN, $L_{\rm{[OIII]}}$  traces the time averaged AGN luminosity over $>100$ years rather than the instantaneous luminosity. Furthermore, $L_{\rm{[OIII]}}$ can suffer from contamination by star formation. All these factors will contribute to the large scatter in the $L_{\rm{[OIII]}}-L_{\rm{bol}}$ relation and make $L_{\rm{[OIII]}}$ only an approximate bolometric luminosity indicator for individual objects.

We note that our best-fit $L_{\rm{[OIII]}}-L_{\rm{bol}}$ relation is consistent with a slope of unity and a bolometric correction of $L_{\rm{bol}}/L_{\rm{[OIII]}}=2200$ (median value). This is somewhat lower than the commonly adopted value for observed [\ion{O}{3}] luminosity of $\sim3500$ \citep{Heckman:2004} derived for local type-1 AGN via the correlation to $L_{5000}$ and the bolometric correction by \citet{Marconi:2004}. We show this value as the reference relation by the black solid line in the $L_{\rm{[OIII]}}-L_{\rm{bol}}$ panel. \citet{Pennell:2017} found a similar average value of 3400 for their sample of type-1 AGN at $0.03<z<1.0$, based on directly integrated SEDs. But they also report a much flatter slope of $0.56$, derived over a similar luminosity range as our study. 
On the other hand, we find that the value $L_{5100}/L_{\rm{[OIII]}}\approx320$ given in \citet{Heckman:2004}, shown as black solid line in the $L_{5100}-L_{\rm{[OIII]}}$ panel in Figure~\ref{fig:Lrelation}, is in perfect agreement with our best-fit relation to the combined FMOS and \citet{Shen:2012b} sample. We thus suspect that the difference stems from their assumed bolometric correction BC$_{5100}=10.9$ compared to BC$_{5100}=7$ adopted in this work. Using the latter value would lead to $L_{\rm{bol}}/L_{\rm{[OIII]}}=2240$ (shown as dotted line in Figure~\ref{fig:Lrelation}), in good agreement with our result. Given the good agreement of our adopted value for BC$_{5100}$ with the bolometric luminosities from \citet{Lusso:2012}, we recommend a value of  $L_{\rm{bol}}/L_{\rm{[OIII]}}\approx2200$ or the relation given in Table~\ref{tab:lumcor} for the bolometric correction of $L_{\rm{[OIII]}}$, if these luminosities are not corrected for intrinsic extinction. 
We note that while we find a clear correlation of $L_{\rm{[OIII]}}$ with both $L_{5100}$ and $L_{\rm{H}\alpha}$, the scatter is increased compared to the correlation with $L_{\rm{bol}}$ (see Table~\ref{tab:lumcor}). While the samples and sample sizes are different, this might indicate that the $L_{\rm{[OIII]}}-L_{\rm{bol}}$ relation is in fact the most fundamental of these and the latter two arise as secondary correlations via $L_{\rm{bol}}$. Interestingly the $L_{5100}-L_{\rm{[OIII]}}$ shows the largest scatter among our correlations, even though both quantities are measured from the same observation, i.e. they suffer least from potential systematics.

We now move our discussion to the correlations with the hard X-ray luminosity \Lx.
The X-ray emission is emitted much closer to the central black hole than the narrow emission lines. The UV photons emitted from the accretion disk are inverse Compton scattered in a plasma of hot electrons known as the corona to X-ray energies. Assuming a universal physical relation between the disk emission and the reprocessed X-ray emission from the hot corona, a tight correlation between bolometric luminosity (or UV  luminosity) and X-ray luminosity would be expected. Indeed such a correlation has been observed and is well studied both for the bolometric luminosity \citep[e.g.][]{Vasudevan:2009,Lusso:2012} and for UV luminosity \citep[e.g.][]{Steffen:2006,Young:2010,Lusso:2016}, although with a typical dispersion of $0.35-0.4$~dex. We confirm the correlation of $L_{\rm{bol}}$ with the intrinsic \Lx\ for our sample, at a dispersion of 0.34~dex. Our data is in good agreement with the commonly adopted luminosity dependent bolometric correction by \citet{Marconi:2004}, shown as black solid line in the $\Lx-L_{\rm{bol}}$ panel in Figure~\ref{fig:Lrelation}. Our best-fit relation is in perfect agreement with \citet{Marconi:2004} at lower luminosities ($\Lx\approx10^{44}$), but slightly deviates at the higher luminosity range ($\Lx\approx10^{45}$).

The second row in Figure~\ref{fig:Lrelation} shows the correlation of \Lx\ with the optical luminosity indicators. While a significant correlation is present in each case, all of the correlations of the optical luminosity indicators with \Lx\ show significant scatter. For the reference relations shown as black lines in each panel we adopt the \citet{Marconi:2004} bolometric correction and BC$_{5100}=7$. For $L_{\rm{H}\alpha}$,  we fold in equation~\ref{eq:l_ab} and for $L_{\rm{[OIII]}}$ we use $L_{5100}/L_{\rm{[OIII]}}=320$. For $L_{5100}$, we find our best-fit relation in good agreement with the reference relation over the luminosity range covered, but slightly steeper when extrapolated beyond that. For  $L_{\rm{H}\alpha}$, our best-fit relation is also in fair agreement with the reference over the luminosity range for the bulk of our sample, but in general it is considerably steeper. It is consistent with unity, rather than the flatter relation predicted by propagating through the luminosity correlations discussed above. Robustly testing the validity of the reference relation requires extending the dynamical range in luminosity of the sample, but is beyond of the scope of the current work.

For $\Lx-L_{\rm{[OIII]}}$ our best-fit it in excellent agreement with the adopted reference relation. In addition we show the empirical relation by \citet{Ueda:2015} as magenta dotted line (their best-fit relation for type~1 AGN, i.e. $N_{\rm{H}}<10^{22}$ cm$^{-2}$, and uncorrected $L_{\rm{[OIII]}}$) and their type-1 AGN subsample as magenta stars. This relation has been established for low-redshift hard X-ray ($>10$~keV) selected AGN from the \textit{Swift} BAT all sky survey  \citep[see also][]{Berney:2015}. While their best-fit relation is in excellent agreement with both our best-fit and our adopted reference relation at low luminosities, it significantly deviates at high luminosities. However, the average $\log \Lx$ of their type-1 subsample is 43.36, significantly fainter than our FMOS sample. Thus the apparent disagreement at high luminosities is caused by extrapolation of their best-fit beyond the main luminosity range of their sample. Indeed the data points of the \textit{Swift} BAT type-1 subsample are fully consistent with our sample. We therefore advocate the use of our best-fit for the $\Lx-L_{\rm{[OIII]}}$ relation for type-1 AGN with uncorrected $L_{\rm{[OIII]}}$, especially at high luminosities, and caution against extrapolating the regression lines by \citet{Ueda:2015} to high luminosities.

We confirm the presence of a non-linear correlation between $\Lx$ and $L_{\rm{[OIII]}}$ suggested by \citet{Ueda:2015} for \textit{Swift} BAT also for the FMOS sample studied here, though with an even flatter slope than found in their study. We argue that the observed non-linearity is fully consistent with the luminosity dependence of the \Lx\ bolometric correction, and thus with the luminosity dependence of the number of X-ray photons to the number of ionizing UV photons emitted by the AGN, as demonstrated by the good agreement with the reference relation which only assumes the luminosity dependent bolometric correction by \citet{Marconi:2004} and constant scale factors to $L_{5100}$ and then to $L_{\rm{[OIII]}}$.

For our sample we find the scatter in the $\Lx-L_{\rm{[OIII]}}$ relation to be comparable or even smaller than for the relation of \Lx\ with $L_{5100}$ and $L_{\rm{H}\alpha}$ and even smaller than for the correlation with $L_{\rm{bol}}$. Given the much larger scatter of $\sim0.5-0.6$~dex found for local AGN \citep{Heckman:2005,Ueda:2015,Berney:2015}, it is not clear if this indicates a tighter physical correlation or is due to sample selection effects. Given the considerable scatter of both  $\Lx$ and $L_{\rm{[OIII]}}$ with $L_{\rm{bol}}$ and the assumption that the physical origin of the scatter for both luminosities are uncorrelated, given the very different spatial scales and physical conditions of their respective emission regions, the comparatively small scatter we find in the $\Lx-L_{\rm{[OIII]}}$ is at least somewhat surprising.

We conclude that we find a consistent set of correlations for various bolometric luminosity indicators for our data set when tied to the bolometric corrections given by \citet{Marconi:2004}, folding in optical luminosity correlations established at lower redshift. We emphasize that we find this good agreement with a few simple assumptions, established independently from our data set at low redshift, namely the bolometric correction from \Lx\ to $L_{\rm{bol}}$ from \citet{Marconi:2004} a bolometric correction to optical continuum luminosity $L_{\rm{bol}}/L_{5100}=7$, a ratio $L_{5100}/L_{\rm{[OIII]}}=320$ and a correlation between $L_{5100}$ and $L_{\rm{H}\alpha}$ given by equation~\ref{eq:l_ab}. This indicates that the basic physics that govern these relationships did not drastically change between $z\sim2$ and $z\sim0$.




\begin{deluxetable*}{lcccc}
\tabletypesize{\scriptsize}
\tablecaption{Spectral measurements for the FMOS \ion{Mg}{2} sample}
\tablewidth{16cm}
\tablehead{
\colhead{Name} & \colhead{CID 113} & \colhead{CID 343}  & \colhead{CID 3576 } & \colhead{CID 781}  
}
\startdata
ID I09 & 1463661 & 797841 & 632431 & 1226535\\  \noalign{\smallskip}
ID XMM & 180 & 146 & 5023 &  \\  \noalign{\smallskip}
RA (J2000) & 150.209 & 149.910 & 149.650 & 150.101\\  \noalign{\smallskip}
DEC (J2000) &  2.482 &  2.081 &  1.866 &  2.419\\  \noalign{\smallskip}
$z_\mathrm{cat}$ &  3.333 &  2.802 &  2.935 &  4.660\\ \noalign{\smallskip}
$z_\mathrm{NIR}$ &  3.358 &  2.809 &  2.936 &  4.643\\ \noalign{\smallskip}
$i$~mag & 20.220 & 20.280 & 19.730 & 22.600\\ \noalign{\smallskip}
$H$~mag & 19.592 & 19.458 & 19.069 & 21.236\\ \noalign{\smallskip}
$J$~mag & 19.484 & 19.545 & 19.142 & 21.892\\ \noalign{\smallskip}
$\log \Lx$ & 44.374 & 44.307 & 44.330 & 44.345\\ \noalign{\smallskip}
S/N$_\mathrm{MgII}$ & 11.582 &  3.237 &  6.082 &  0.789\\ \noalign{\smallskip}
S/N$_\mathrm{MgII,Line}$  &  7.889 &  3.732 &  4.980 &  3.768\\ \noalign{\smallskip}
$\log L_\mathrm{MgII}$ & $44.031 \pm  0.020$ & $44.048 \pm  0.056$ & $43.981 \pm  0.042$ & $43.857 \pm  0.068$\\ \noalign{\smallskip}
FWHM$_\mathrm{MgII}$  & $3437 \pm 201$ & $3422 \pm 568$ & $1745 \pm 249$ & $1929 \pm 295$\\  \noalign{\smallskip}
$\log L_{3000}$ & $46.061 \pm  0.012$ & $46.011 \pm  0.032$ & $45.919 \pm  0.036$ & $45.377 \pm  0.138$\\  \noalign{\smallskip}
$\log \mbh$ &  $ 9.090 \pm  0.056$ & $ 9.055 \pm  0.150$ & $ 8.413 \pm  0.134$ & $ 8.164 \pm  0.157$\\  \noalign{\smallskip}
$\log L_\mathrm{bol}$ & $46.565 \pm  0.012$ & $46.517 \pm  0.032$ & $46.428 \pm  0.036$ & $45.926 \pm  0.138$\\  \noalign{\smallskip}
$\log \er$ & $-0.635 \pm  0.055$ & $-0.648 \pm  0.146$ & $-0.095 \pm  0.134$ & $-0.348 \pm  0.149$\\ \noalign{\smallskip}
$t_\mathrm{growth}$ [Myr] &  224.6 &  230.2 &  56.3 &  95.0\\ \noalign{\smallskip}
$t_\mathrm{growth}/t_\mathrm{uni}$&  1.197 &  1.011 &  0.260 &  0.754 \\
\enddata
\tablecomments{Sample properties and spectral measurements for the four AGN at $z>2.7$ with broad \ion{Mg}{2} detection. All of them have been observed in COSMOS in LR mode. ID~I09 is the optical counterpart ID in the catalog by \citet{Ilbert:2009}, ID~XMM is the ID in XMM-COSMOS, $z_\mathrm{cat}$ and  $z_\mathrm{NIR}$ denote the redshift reported in \citet{Marchesi:2016} and as measured from the \ion{Mg}{2}}
\label{tab:mgii}
\end{deluxetable*}

\begin{figure}
\centering
\resizebox{\hsize}{!}{ \includegraphics[clip]{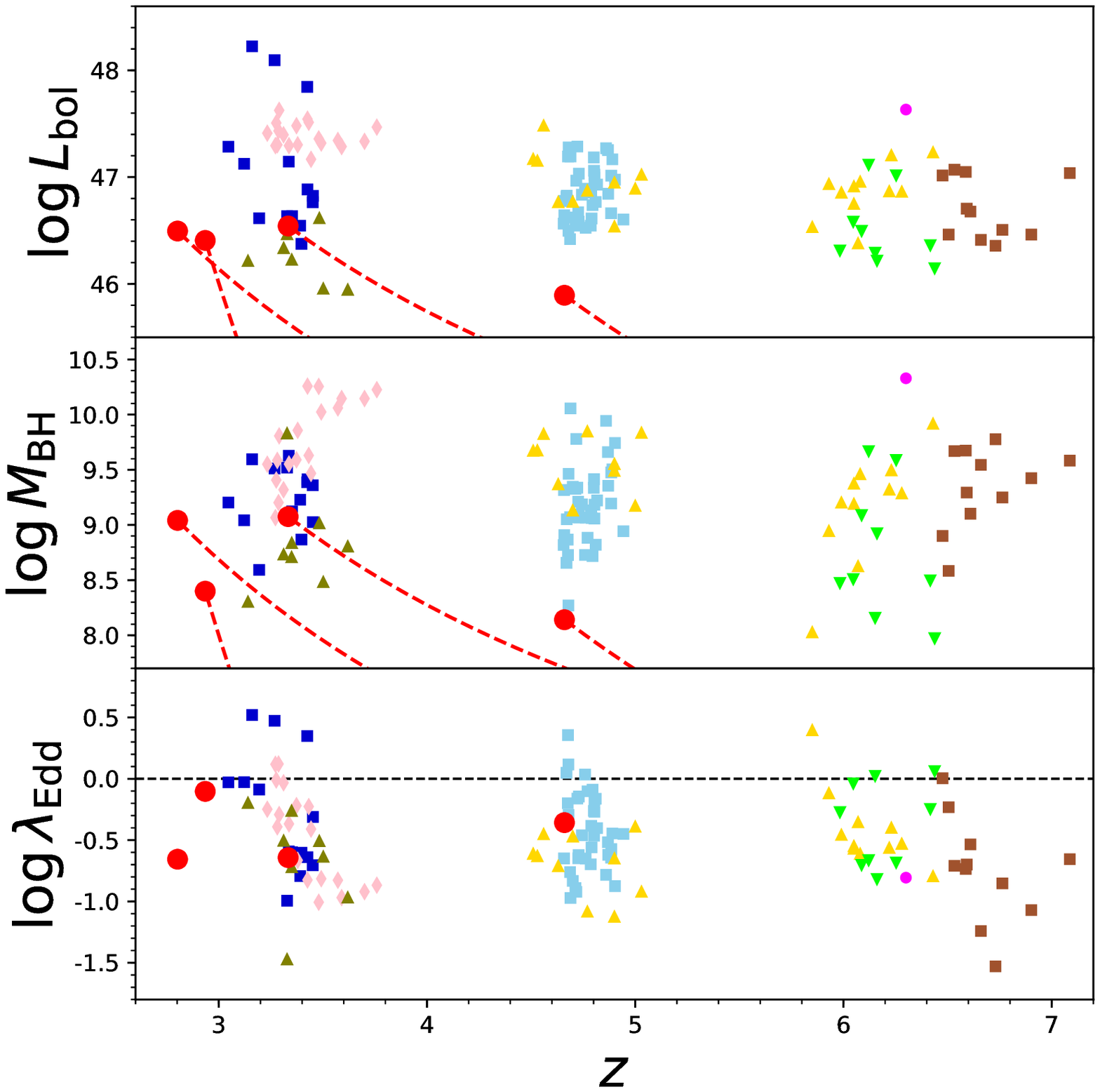} }
\caption{Overview of the distribution of $L_{\mathrm{bol}}$ (upper panel), \mbh\  (middle panel) and \er\ (lower panel) for near-IR observations of broad-line AGN samples at $2.7<z<7.1$. The four FMOS AGN are marked as large red circles. The red dashed line shows their potential growth path predicted for the toy model discussed in the text. The other symbols represent samples from the literature with reliable \mbh\ estimates based on either H$\beta$ or \ion{Mg}{2}.  At $z\sim3.5$, these are the AGN are from \citet{Shemmer:2004}, \citet[][blue squares]{Netzer:2007}, \citet[][pink diamonds]{Zuo:2015} and \citet[][olive triangles]{Trakhtenbrot:2016}. At $z\sim4.8$, these are the AGN samples are from \citet[][cyan squares]{Trakhtenbrot:2011} and \citet[][gold triangles]{DeRosa:2011}, based on \citet{Iwamuro:2002}, and at $z\gtrsim6$ again \citet[][gold triangles]{DeRosa:2011}, \citet[][green downward triangles]{Willott:2010}, \citet[][magenta circle]{Wu:2015} and \citet[][brown squares]{Mazzucchelli:2017}.  The horizontal dashed line in the lower panel indicates the Eddington limit.}
\label{fig:highz} 
\end{figure}

\subsection{Black hole growth at $z\gtrsim3$}
In this section, we focus on the small but  interesting sub-sample of moderate-luminosity AGN at $z\gtrsim3$ for which we detect a broad \ion{Mg}{2} line. In total, we obtained \ion{Mg}{2} based \mbh for four AGN, including our highest-$z$ source CID~781 at $z=4.64$. We list their properties and spectral measurements in Table~\ref{tab:mgii}.
In Figure~\ref{fig:highz}, we show their distribution in $L_{\mathrm{bol}}$, \mbh\ and \er, as a function of redshift, together with samples from the literature at $z>2.5$ with reliable black hole mass estimates from either \ion{Mg}{2} or H$\beta$. These include samples of luminous QSOs at $z\sim3.5$ \citep{Shemmer:2004,Netzer:2007,Zuo:2015}, at $z\sim4.8$ \citep{Iwamuro:2002,Trakhtenbrot:2011} and $z\sim6.2$ \citep{Willott:2010,DeRosa:2011,Wu:2015,Mazzucchelli:2017} as well as of moderate-luminosity AGN at  $z\sim3.5$ in COSMOS from \citet{Trakhtenbrot:2016}.  We note that one AGN in our sample (CID~113) is also included in the latter study. While we observe \ion{Mg}{2} in J-band, \citet{Trakhtenbrot:2016} measured \mbh from broad H$\beta$ observed in K-band. Our \mbh estimate of $10^{9.1} M_\odot$ is consistent with their H$\beta$ based \mbh estimate of $10^{8.78} M_\odot$ within the uncertainties of the virial method.

Our three AGN at $z\sim3$ fall in a similar parameter space as those from previous studies \citep{Netzer:2007,Trakhtenbrot:2016}. CID~781 on the other hand occupies a luminosity regime which has not been probed yet at $z\sim4.7$, being $>0.5$~dex fainter than previous near-IR studies \citep{Trakhtenbrot:2011}. Its black hole mass, of about $10^8\,M_\odot$, is at the low end of the currently probed \mbh distribution and about 1~dex below the typical masses of more luminous quasars at the same redshift. The Eddington ratio however is similar to more luminous quasars at high-$z$. 
It can serve as an interesting analog to low-luminosity quasars recently discovered in large numbers at $z\gtrsim6$ by HSC-SSP \citep{Matsuoka:2016,Matsuoka:2018,Matsuoka:2018b}.

We note that all our high$-z$ AGN have fairly high \er\ ($\log \er>-0.7$), while our survey would be sensitive to detect lower \er\ objects with $\mbh>10^{9} M_\odot$. This is consistent with the results from \citet{Trakhtenbrot:2016} and with observations for $z\gtrsim6$ quasars \citep[e.g.][]{Willott:2010}, while also rare cases with low \er\ have been discovered at these redshifts \citep{Trakhtenbrot:2015,Kim:2018}. This might indicate on average higher \er\ for the $z\gtrsim3$ AGN population than what is found at lower redshift \citep{Schulze:2015}, i.e. a redshift evolution in the shape of the ERDF. Robustly confirming this suggestion requires a full determination of the underlying active BHMF and ERDF at $z\gtrsim3$, based on black hole mass estimates for a well-defined AGN sample (Schulze et al., in prep).
 
We next use our  $z\gtrsim3$  AGN sample to evaluate their constraints on the black hole growth in the early universe. 
While black hole growth is most likely a stochastic process with significant variation in their accretion rates over time scales of $>10^{5}$~years \citep[e.g.][]{Novak:2011,Schawinski:2015}, we here explore a simplified growth model in which we assume accretion at a constant Eddington ratio and with constant radiative efficiency \citep[e.g.][]{Salpeter:1964,Trakhtenbrot:2011}. This case corresponds to an exponential growth with an e-folding time of 
\begin{equation}
\tau = 4 \times 10^8 \frac{\eta}{\er (1-\eta)} \ \rm{yr}  \label{eq:tgrowth} 
\end{equation}
and a growth time of $t_{\rm{growth}}=\tau\ln \left( \mbh / M_{\rm{seed}} \right)$.
For simplicity, we here assume $\eta=0.1$, $M_{\rm{seed}}=10^4 M_\odot$ and constant accretion at the measured \er, consistent with previous studies \citep{Netzer:2007b,Trakhtenbrot:2011}. We show the growth path for our $z\gtrsim3$ AGN under this simplified scenario as red dashed lines in Figure~\ref{fig:highz} and provide the growth times in Gyr and normalized by the age of the universe at the respective $z$ in Table~\ref{tab:mgii}. Their currently high \er correspond to $e$-folding timescales of $56-230$~Myr. For the two higher \er\ sources CID~781 and CID~3576 these timescales are fast enough to allow the SMBHs to grow to their current masses under our simplistic scenario. For the other two, CID~113 and CID~343, their inferred growth times are around or slightly exceed the age of the universe at their redshift. This suggests that they most likely have been growing at a higher rate at least for some fraction of the time, since we made the optimistic assumption of continuous growth, i.e. a duty cycle of unity. Alternative solutions involve a higher seed black hole mass or a lower radiative efficiency as assumed here. 
We conclude that with our sample we are probing a more typical \mbh regime, $10^8-10^9 M_\odot$, in an evolutionary  phase of fast growth at $z\gtrsim3$.

\section{Conclusions} \label{sec:conclu}
We present near-IR spectroscopy in $J$- and $H$-band for a large sample of 243 X-ray selected moderate-luminosity type-1 AGN in the extragalactic survey fields of COSMOS, SXDS and E-CDF-S using the multi-object spectrograph Subaru/FMOS. Our sample covers the redshift range $0.5<z<4.7$ (with the vast majority at $z<2.6$) over an X-ray luminosity range of $10^{43}\lesssim \Lx \lesssim 10^{45}$~erg s$^{-1}$. Broad H$\alpha$ is detected in 211 AGN, broad H$\beta$ in 63 and \ion{Mg}{2} is covered in the FMOS spectra in 4 AGN at $z>2.7$. We fit parametric models to the near-IR spectra, measure line widths, line and continuum luminosities and estimate black hole masses for our targets. We supplement these with \ion{Mg}{2}  and \ion{C}{4} measurements from optical spectra and compare line widths, luminosities and black hole mass estimates using different broad emission lines. In addition, we generate composite spectra for our FMOS sample and compare them with composite spectra of more luminous AGN at similar redshift and with low-$z$ AGN at matched luminosity. Our main results are the following.
\begin{enumerate}
\item We provide a catalog of estimates of \mbh, $L_{\rm{bol}}$ and \er for a sample of 243 AGN in the deep fields COSMOS, SXDS and E-CDF-S, with their rich multi-wavelength coverage. Our results enhance the legacy value on AGN studies in these fields and enable future studies on SMBH-galaxy co-evolution, AGN physics and others. \rev{We make the catalog available online.\footnote{\href{http://member.ipmu.jp/fmos-cosmos/fmosBL.fits}{http://member.ipmu.jp/fmos-cosmos/fmosBL.fits}.}}
\item We confirm the validity of several correlations from the literature between broad-line FWHM, broad-line luminosity and continuum luminosity between H$\alpha$, H$\beta$, \ion{Mg}{2} and \ion{C}{4} for our data set.
\item We show that black hole mass estimates from H$\alpha$, H$\beta$ and \ion{Mg}{2} are unbiased and consistent with each other on average for our moderate-luminosity AGN sample when consistently calibrated virial mass formulas are used. There is a non-negligible amount of scatter between our \mbh\ estimates from different lines, due to a combination of limited data quality and intrinsic effects.
\item For the  \ion{C}{4} line, we find a considerable scatter in the comparison of their FWHM to those from H$\alpha$ and H$\beta$ and consequently also in the \ion{C}{4} based \mbh estimates, confirming the large uncertainties associated to using \ion{C}{4} as virial \mbh estimator. Using a correction based on the \ion{C}{4} blueshift leads to an improvement in the \mbh estimates, but for our moderate-luminosity AGN sample such a correction is of less importance than for more luminous AGN.
\item We find differences in the composite spectra between our moderate-luminosity sample and a high luminosity sample in line with expectations given their different bolometric luminosities, $M_{\rm{BH}}$, and \er. We also directly confirm the presence of a Baldwin effect like trend in the [\ion{O}{3}] line at $z>1$ and an enhanced prominence of ionized outflow indications in luminous AGN compared to our moderate-luminosity AGN sample.
Comparison with a lower redshift, luminosity matched sample shows good agreement for the H$\beta$ region, but we find on average lower H$\alpha$ EW in the higher-$z$ FMOS sample.
\item The observed luminosity correlations between $L_{\rm{bol}}$, \Lx, $L_{\rm{[OIII]}}$, $L_{5100}$ and $L_{\rm{H}\alpha}$ with each other are consistent with a simple empirical model based on the  $L_{\rm{bol}}- \Lx$ bolometric correction by \citet{Marconi:2004}, $L_{\rm{bol}}/L_{5100}=7$, $L_{5100}/L_{\rm{[OIII]}}=320$ and a correlation between $L_{5100}$ and $L_{\rm{H}\alpha}$ given by Equation~\ref{eq:l_ab}. 
\item We have detected broad \ion{Mg}{2} in CID~781, a moderate-luminosity AGN at $z\sim4.6$, and three additional AGN at  $z\sim3$. CID~781 occupies a luminosity regime not yet probed before at $z>4$ with near-IR spectroscopy. Its current growth rate is fast enough to allow growth to its current mass from a $10^4 M_\odot$ seed black hole. For the $z\sim3$ objects 2/3 require growth at a faster rate than suggested by their current accretion rate at least over some periods in their past.
\end{enumerate}

\acknowledgements
We thank the anonymous referee for the constructive report which improved the clarity of the paper.
A.S. is supported by the EACOA fellowship and acknowledges support by JSPS KAKENHI Grant Number 26800098. This work was supported by JSPS KAKENHI Grant Number 26400221 (PI J. Silverman) and the World Premier International Research Center Initiative (WPI), MEXT, Japan.
N.A. is supported by the Brain Pool Program, which is funded by the Ministry of Science and ICT through the National Research Foundation of Korea (2018H1D3A2000902).

Based on data products from observations made with ESO Telescopes at the La Silla Paranal Observatory as part of the VISTA Deep Extragalactic Observations (VIDEO) survey, under programme ID 179.A-­2006 (PI: Jarvis) and  under ESO programme ID 179.A-2005 and on data products produced by TERAPIX and the Cambridge Astronomy Survey Unit on behalf of the UltraVISTA consortium.


%
%

%

\bibliographystyle{aasjournal}  
\bibliography{ref_fmos}

\end{document}